\documentclass[aps,prc,twocolumn,groupedaddress,nofootinbib]{revtex4-1}
\usepackage{rotating}% Include figure files
\usepackage{graphicx}% Include figure files
\usepackage{dcolumn}% Align table columns on decimal point
\usepackage{bm}% bold math
\usepackage{sidecap}
\usepackage{xspace}
\usepackage{multirow} 
\usepackage{amsmath}
\usepackage{makecell}

\newcommand{\beq}{\begin{equation}}
\newcommand{\eeq}{\end{equation}}
\newcommand{\beqa}{\begin{eqnarray}}
\newcommand{\eeqa}{\end{eqnarray}}
\newcommand{\snn}{\sqrt{s_{_{NN}}}}

\newcommand{\npart}{N_{\rm part}}

\newcommand{\pT}{$p_{T}$}

\newcommand{\mylessthan}{$<$}
\newcommand{\Erel}{\mbox{$E_{\rm rel}$}\xspace}

\begin{document}

\title{Participant and spectator scaling of spectator fragments in Au+Au and Cu+Cu\\
collisions at $\sqrt{s_{NN}}$\,=\,19.6~and~22.4\,GeV.}
\author{B.Alver$^4$,
B.B.Back$^1$,
M.D.Baker$^2$,
M.Ballintijn$^4$,
D.S.Barton$^2$,
R.R.Betts$^6$,
A.A.Bickley$^7$,
R.Bindel$^7$,
A.Budzanowski$^3$,
W.Busza$^4$,
A.Carroll$^2$,
Z.Chai$^2$,
V.Chetluru$^6$,
M.P.Decowski$^4$,
E.Garc\'{\i}a$^6$,
T.Gburek$^3$,
N.George$^{1,2}$,
K.Gulbrandsen$^4$,
S.Gushue$^2$,
C.Halliwell$^6$,
J.Hamblen$^8$,
I.Harnarine$^6$,
G.A.Heintzelman$^2$,
C.Henderson$^4$,
D.J.Hofman$^6$,
R.S.Hollis$^6$,
R.Ho\l y\'{n}ski$^3$,
B.Holzman$^2$,
A.Iordanova$^6$,
E.Johnson$^8$,
J.L.Kane$^4$,
J.Katzy$^{4,6}$,
N.Khan$^8$,
W.Kucewicz$^6$,
P.Kulinich$^4$,
C.M.Kuo$^5$,
W.Li$^4$,
W.T.Lin$^5$,
C.Loizides$^4$,
S.Manly$^8$,
D.McLeod$^6$,
A.C.Mignerey$^7$,
R.Nouicer$^6$,
A.Olszewski$^3$,
R.Pak$^2$,
I.C.Park$^8$,
H.Pernegger$^4$,
C.Reed$^4$,
L.P.Remsberg$^2$,
M.Reuter$^6$,
E.Richardson$^7$,
C.Roland$^4$,
G.Roland$^4$,
L.Rosenberg$^4$,
J.Sagerer$^6$,
P.Sarin$^4$,
P.Sawicki$^3$,
I.Sedykh$^2$,
W.Skulski$^8$,
C.E.Smith$^6$,
M.A.Stankiewicz$^2$,
P.Steinberg$^2$,
G.S.F.Stephans$^4$,
A.Sukhanov$^2$,
A.Szostak$^2$,
J.-L.Tang$^5$,
M.B.Tonjes$^7$,
A.Trzupek$^3$,
C.Vale$^4$,
G.J.van~Nieuwenhuizen$^4$,
S.S.Vaurynovich$^4$,
R.Verdier$^4$,
G.I.Veres$^4$,
P.Walters$^8$,
E.Wenger$^4$,
D.Willhelm$^7$,
F.L.H.Wolfs$^8$,
B.Wosiek$^3$,
K.Wo\'{z}niak$^3$,
A.H.Wuosmaa$^1$,
S.Wyngaardt$^2$,
B.Wys\l ouch$^4$\\
\vspace{3mm}
\small
$^1$~Physics Division, Argonne National Laboratory, Argonne, IL 60439-4843, USA\\
$^2$~Brookhaven National Laboratory, Upton, NY 11973-5000, USA\\
$^3$~Institute of Nuclear Physics, Krak\'{o}w, Poland\\
$^4$~Laboratory for Nuclear Science, Massachusetts Institute of Technology, Cambridge, MA 02139-4307, USA\\
$^5$~Department of Physics, National Central University, Chung-Li, Taiwan\\
$^6$~Department of Physics, University of Illinois at Chicago, Chicago, IL 60607-7059, USA\\
$^7$~Department of Chemistry, University of Maryland, College Park, MD 20742, USA\\
$^8$~Department of Physics and Astronomy, University of Rochester, Rochester, NY 14627, USA\\
}
\date{\today}

\begin{abstract}
Spectator fragments resulting from relativistic heavy
ion collisions, consisting of single protons and
neutrons along with groups of stable nuclear
fragments up to Nitrogen ($Z$\,=\,7), are measured in PHOBOS.
These fragments are observed in Au+Au ($\snn$\,=\,19.6\,GeV)
and Cu+Cu (22.4\,GeV) collisions at high pseudorapidity ($\eta$).
The dominant multiply-charged fragment is the
tightly bound Helium ($\alpha$), with Lithium, Beryllium, and Boron
all clearly seen as a function of collision centrality and
pseudorapidity.
We observe that in Cu+Cu collisions, it becomes much more
favorable for the $\alpha$ fragments to be released than Lithium.
The yields of fragments approximately scale with the number of
spectator nucleons, independent of the colliding ion.
The shapes of the pseudorapidity distributions of fragments indicate that the
average deflection of the fragments away from the beam direction
increases for more central collisions.
A detailed comparison of the shapes for $\alpha$ and Lithium fragments
indicates that the centrality dependence of the deflections favors a 
scaling with the number of participants in the collision.
\end{abstract}


\maketitle

\section{Introduction}
In relativistic heavy ion collisions, the nucleons of the interacting ions
can be divided into two distinct categories:
those that experience an inelastic collision with at
least one nucleon from the opposing nucleus (participants)
and those that do not (spectators).
Participant nucleons ultimately create the bulk of
particles observed in the detectors. Spectators consist of
single protons and neutrons as well as
larger spectator fragments including Helium, Lithium, 
Beryllium, Boron, and higher mass nuclei.
Na\"{\i}vely, these spectators are free to continue along
their original path as they do not directly participate in
the collision.  In practice, however, they can interact
in several ways and still be considered a spectator by
the usual definition: for example they can suffer an elastic collision
with a nucleon from the other beam, they can be affected
by any remaining nuclear binding energy in the beam remnant,
or they can interact with produced particles from the
participant zone~\cite{cite:PHOBOS_v1}. 

Fragmentation of nuclei has been studied in a number of
experiments~\cite{cite:EMU,cite:SinghJain,cite:KLMM_PhysRevC,cite:ALADIN,cite:KLMM_ZPhys,cite:KLM_ActaPhysPol,cite:E877,cite:DWW}.
These experiments typically covered the full kinematic and solid
angle range needed to accurately identify all fragments and
basic fragment properties such as $A$ and $Z$, and their momenta.
However, these experiments suffered from a lack of statistics,
with only $\mathcal{O}(1000)$ events in total, precluding detailed
differential studies of fragmentation properties as a function of
impact parameter.

The observed properties of fragments, such as their momentum vectors, can be
described by a combination of the beam momentum at the time of
the collision and the internal Fermi motion within the nucleus
in its rest frame.
In the absence of Fermi motion and other external effects,
spectator fragment transverse momenta would be zero and they would
consequently continue traveling at the same rapidity as the beam.
In this limit, the polar angle ($\theta$) of fragments would be zero
or, equivalently, they would have infinite pseudorapidity ($\eta$):

\begin{equation}
\label{eqn:eta_theta}
\begin{split}
\ensuremath{\eta & = -{\rm ln}({\rm tan}(\theta/2))}\\
\ensuremath{     & \rightarrow \infty (\theta\rightarrow0).}
\end{split}
\end{equation}

\noindent Including the Fermi motion, however, leads to a finite
transverse momentum component of the fragments and reduces the
particle rapidity to below that of the beam.  With a finite
(nonzero) polar angle, it is possible that the products will be
intercepted by active elements of a detector.
In addition, the internal Fermi motion also modifies the
longitudinal component of the momentum, however this effect
is typically small compared to the boosted momentum of the nucleons.

Transverse momentum is boost invariant and it therefore becomes useful
to compare data across multiple experiments with differing
collision energies.
Equivalently, by converting the momentum vectors into an
angular form, one can show that the pseudorapidity density
distribution ($dN/d\eta$ versus $\eta$) becomes approximately
boost invariant, which also allows for the comparison of data at
different $\snn$.  To account for energy differences, one subtracts
the rapidity of the beam at the appropriate energy scale; 
a nontrivial transformation described in Appendix~\ref{ap:EtaPrimer}.

In the PHOBOS experiment~\cite{cite:PHOBOS_NIM}
at the Relativistic Heavy-Ion Collider
(RHIC), completely-freed neutrons can be measured using the
Zero-Degree Calorimeters (ZDC)~\cite{cite:ZDCs}, which are specifically designed
for this purpose. Charged fragments are not
observed in these detectors as they are swept away from
the ZDCs by the RHIC accelerator magnets. A calorimeter that could
detect very forward protons was available for some PHOBOS running
periods, but was not used in this analysis.
At RHIC injection energies, nucleon-nucleon center of mass energy
$\snn$\,=\,19.6~(Au+Au) and~22.4\,GeV (Cu+Cu), spectators with a 
finite transverse momentum can be detected 
within the pseudorapidity acceptance of PHOBOS.  However, the
finite acceptance of the detector limits the measurement of very
low-$p_T$ particles, especially for large-$Z$ fragments.
A large statistical sample, though, has been amassed which does allow
for some more detailed studies not afforded to other experiments.

This paper presents detailed measurements of large-$Z$ fragments in the
PHOBOS detector.  Section~\ref{sec:Detector} describes the detector.
Section~\ref{sec:Analysis} describes the analysis
methods used to distinguish differently charge particles.
Sections~\ref{sec:EtaDep}~and~\ref{sec:CentDep}
show the pseudorapidity and centrality dependencies of the
fragments, respectively.  Section~\ref{sec:EtaCent} discusses how,
in combining the system size, centrality, and pseudorapidity dependencies,
one can probe scaling effects of the large-$Z$ fragments in the 
context of the number of spectators and participants in the collision.

\section{PHOBOS Detector}\label{sec:Detector}
\begin{figure}[!t]
\centering
\includegraphics[angle=0,width=0.475\textwidth]{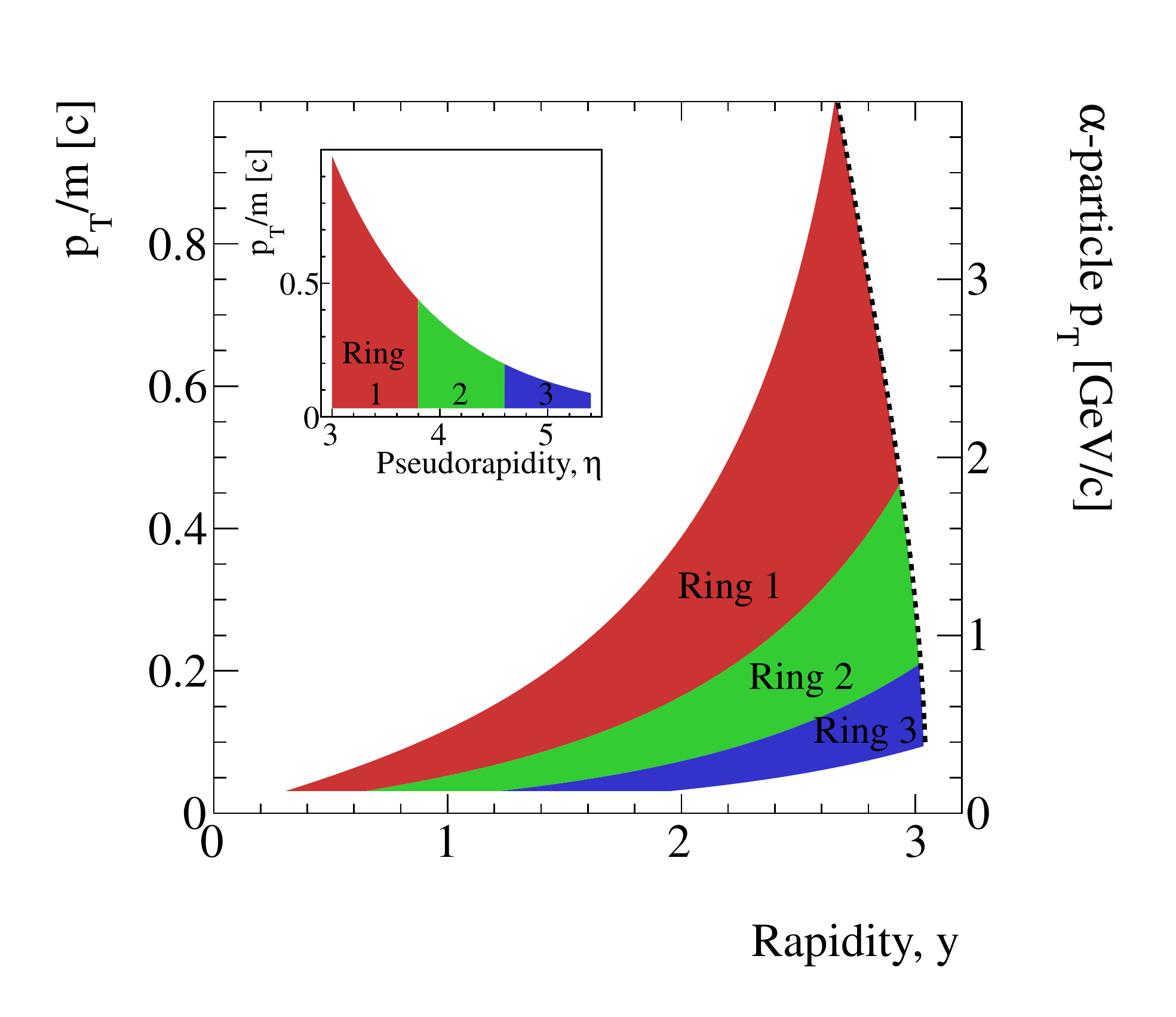}%1
\caption{\label{fig:Acceptance}
(color online) Transverse momentum and rapidity coverage of charged
particles in the silicon Ring detectors in PHOBOS.  The
main figure shows the \pT$/m$-rapidity acceptance for
charged particles in each Ring (different shaded bands).  The
boundary on the rightmost edge of the shaded region
depends on the beam energy.  The dashed line shows the boundary for 
$p_{z}/m$\,=\,$p_{\rm beam}/m_{Au}$ for $\snn$\,=\,19.6\,GeV Au+Au collisions.
The right-hand axis shows the $p_{T}$-scale for $\alpha$ particles, i.e.
$m$\,=\,3.727\,GeV/$c^{2}$.
The inset figure shows the
Ring-detector $p_T$ and pseudorapidity coverage.}
\end{figure}

PHOBOS is a large acceptance silicon detector, covering
almost 2$\pi$ in azimuth and $|\eta|$\mylessthan$5.4$ ($\theta$\,$>$\,$9$\,mrad)~\cite{cite:PHOBOS_NIM}. For the results presented
here, the energy loss measured in the Ring detectors
(3.0\mylessthan$|\eta|$\mylessthan$5.4$) is used to identify spectator
fragments.  The Rings are silicon pad detectors arranged
in an octagonal pattern perpendicular to and surrounding the beam pipe. Three
Ring detectors are placed on each side of the interaction
point at approximately 1,~2,~and~5\,meters from the center of the
interaction region. This
configuration allows for full coverage with minimal
overlapping areas.
In addition, the Octagon silicon barrel, which
consists of a single-layer of silicon parallel to and surrounding the
beam pipe covering $|\eta|$\mylessthan$3.2$, is used for collision vertex
and event centrality determination.

In order to distinguish between singly- and multiply-charged fragments,
the relative energy loss, \Erel,
is defined as

\begin{equation}
\label{eqn:Erel}
\ensuremath{{E_{\rm rel}} = \frac{E_{\rm loss}}{\langle E_{\rm loss} \rangle|_{Z=1} },}
\end{equation}

\noindent where $E_{\rm loss}$ is the energy loss in the silicon detector
and $\langle E_{\rm loss} \rangle|_{Z=1}$ is the mean energy loss for a 
$Z$\,=\,1 particle.  Singly-charged particles (for example spectator protons,
deuterons, and tritons) and singly-charged participants or produced particles
(created by the participants) all appear at an \Erel position close to 1 and, as such,
cannot be separated.
For larger fragments,
with charge greater than unity, energy loss in the silicon
follows a charge-squared ($Z^{2}$) dependence, leading to the
appearance of $\alpha$ particles (for example) at four times the \Erel position of
a singly-charged particle.
%  Similarly, one expects even
%larger fragments at $n^2$ times the nominal \Erel position, for
%an integer $n>1$.

The transverse momentum, $p_{T}$, and rapidity, $y$, coverage for charged
particles in the Rings is shown in Fig.~\ref{fig:Acceptance}.
As there is no significant magnetic field traversed by
forward-going particles, the fixed $\eta$ Ring boundaries
translate to fixed curves in $p_{T}/m$ versus $y$ for
all charged particles. The high-$p_{T}$ and $y$ boundary (rightmost
edge for each Ring) is calculated for $\snn$\,=\,19.6\,GeV Au+Au collisions,
assuming a maximum $p_{z}/m =  p_{\rm beam}/m_{Au}$, where $p_{z}$ is the
momentum of the particle (of mass $m$) along the beam direction,
and $p_{\rm beam}$ is the beam momentum.
 
\section{Data Analysis}\label{sec:Analysis}
\subsection{Event Selection}
The data were recorded during the 2001 (Au+Au -- 
$\snn$\,=\,19.6\,GeV) and 2005 (Cu+Cu --
$\snn$\,=\,22.4\,GeV) RHIC runs. Readout of the silicon was initiated
by a minimally biased trigger for each data set based on coinciding signals from
two arrays of 16 plastic scintillators (3.2\mylessthan$|\eta|$\mylessthan$4.5$), the ``Paddle'' trigger counters~\cite{cite:PHOBOS_PaddleNIM}. For
Au+Au (Cu+Cu) collisions, a minimum of 3~(1) scintillator
hits were required in each array to start readout. The
collision vertex position along the beam line ($z$) was determined via a
probabilistic approach using hits in the Octagon silicon
barrel~\cite{cite:PHOBOS_VertexNIM}.  For Cu+Cu collisions at
$\snn$\,=\,22.4\,GeV, a vertex requirement of $|z|$\mylessthan$10$\,cm
from the nominal vertex position was imposed; for Au+Au
this was relaxed to $|z|$\mylessthan$20$\,cm to maximize the statistics from
the single day-long run.
A total of 84k (2.1M) events were selected for this analysis out of 327k
(15.7M) recorded, respectively for Au+Au (Cu+Cu) collisions.  Events
are dominantly rejected due to the vertex requirement.
The estimated trigger efficiency (coupled with the
vertex finding efficiency)
for the Au+Au (Cu+Cu) data set is 83.5$\pm$3\% (79$\pm$5\%),
determined using the same methods as described in
Ref.~\cite{cite:PHOBOS_200_20} with
the data divided into seven (six) centrality classes,
each with 10\% of the total nuclear inelastic cross-section.
The centrality measure, EOct, is the summed energy loss in the silicon
of the centrally located Octagon barrel in the region $|\eta|$\,$<$\,3.0~\cite{cite:PHOBOS_200_20}.  
The EOct parameter is defined in a $|\eta|$ region smaller than the full acceptance of the Octagon to limit any systematic effects of acceptance shifts
(due to the collision vertex position) and to
reduce the overlap with the Ring detector acceptance.
The lowest centrality cut-off is defined as the point
at which the trigger+vertex efficiency falls below 100\%.
For each
centrality class, the number of participants ($\npart$) is estimated
by use of a Glauber model calculation~\cite{cite:PHOBOS_Glauber}.
Also, the number of spectator nucleons emitted at either the positive or
negative pseudorapidity is calculated as
$N_{\rm spec}/2$\,=\,($\npart^{\rm max}$-$\npart$)/2, where
$\npart^{\rm max}$\,=2$A$\,=\,394~(126) for Au~(Cu) nuclei.

\begin{figure}[!t]
  \centering
  \includegraphics[angle=0,width=0.431\textwidth]{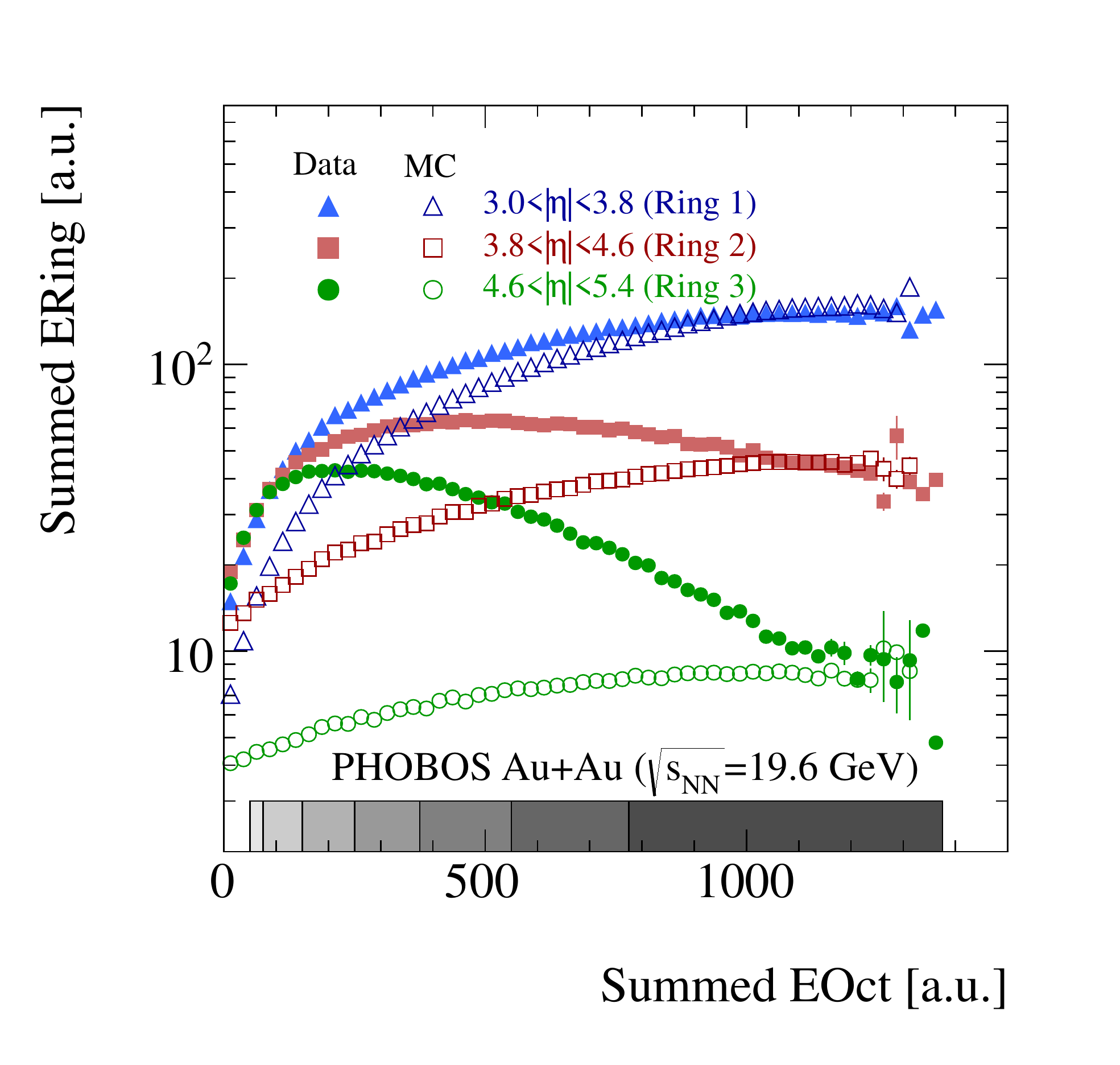}%2
  \caption{\label{fig:ERingVsEOct}
    (color online) Correlation between the summed energy recorded in each of
    the Ring detectors (ERing) and the summed energy deposited in the Octagon
    barrel (EOct) in Au+Au collisions at $\snn$\,=\,19.6\,GeV.
    Filled (open) symbols illustrate the measured distributions from data
    (simulation).  Spectators have been explicitly excluded from the
    simulation distributions.
    The bands show the centrality class selection bins used in this
    analysis, with darker bands corresponding to more central events.
    See text for discussion.}
\end{figure}

\subsection{Motivation}

The first observation of the presence of charged spectator fragments, in the 
acceptance of PHOBOS, was made during the first low-energy
data~\cite{cite:SigFrag}.
The measured charged particle multiplicity
was found to be larger at high pseudorapidity in peripheral
data than in central data, an opposite effect than was expected, and
in contrast to the observed dependencies at mid-rapidity.
Several tests were performed to confirm that the larger
particle yield at high pseudorapidity likely originated from spectator fragments.
Figure~\ref{fig:ERingVsEOct} shows the correlation between
the summed energy in each silicon ring (ERing) and the summed
energy deposited in the silicon Octagon barrel (EOct). Filled symbols
represent data; open symbols show the result of a
Monte-Carlo (MC) simulation that uses particles generated from a
{\sc Hijing}~\cite{cite:HIJING} event simulation passed through
a full {\sc Geant}~\cite{cite:GEANT} description of the PHOBOS
detector and has had spectator fragments
explicitly removed from the acceptance of the detector.

In the MC simulation, a monotonic correlation is observed
between ERing~1 and EOct, which becomes weaker for
larger pseudorapidities.  Even at the highest pseudorapidities, ERing~3
still increases with increasing EOct.
In the data, the dependence of ERing~1 on EOct is
similar in shape to that found in the MC simulation. At higher
pseudorapidities, however, the positive correlation is
restricted to the lowest EOct range and, after reaching a maximum,
ERing~2 and ERing~3 start to decrease with increasing EOct. 

This same anticorrelated dependence was
observed in Au+Au data at higher energies in the correlation
between the Paddle Scintillator counters and the Zero Degree
Calorimeter (ZDCs).  The ZDCs detect spectator neutrons and
include roughly the same relative $\eta$ region (i.e. when considering
the difference in beam rapidities ($y_{\rm beam}$) for different
collision energies: $\eta$\,--\,$y_{\rm beam}$) in
$\snn$\,=\,200\,GeV collisions
as covered by Rings 2 and 3 for 19.6\,GeV, see for
example Ref.~\cite{cite:PHOBOS_130MidRap}.
It is possible that the multiplicity distribution from produced
particles narrows for more central collisions~\cite{cite:BigMultPaper},
however this could not account for the observed rise/fall behavior.

\subsection{Fragment Identification}

Fragments are identified using their relative energy loss (\Erel)
in the silicon (see Eq.~\ref{eqn:Erel}).
Figure~\ref{fig:AuAuMethod} shows the \Erel distribution measured 
in the ERing acceptance for Au+Au collisions at
$\snn$\,=\,19.6\,GeV, where no centrality selection is made
and only the region 5.0$<$$|\eta|$$<$5.4 is shown in order
to make the higher mass fragments more pronounced. 
In Fig.~\ref{fig:AuAuMethod}, the data is shown as
a blue spectrum along with the distribution expected from singly-charged
particles ($Z$\,=\,1, red).  The latter is
considered to be a ``background'' to the data and is
determined from a MC simulation without spectator
fragments.  This $Z$\,=\,1 contribution can be explicitly subtracted
as it is entirely due to singly-charged particles (mostly
from the collision) with a typical Landau-like distribution.

\subsection{Subtracting Singly-Charged Particles}

\begin{figure}[h]
%  \begin{minipage}{0.55\textwidth}
    \centering
    \includegraphics[width=0.99\linewidth]{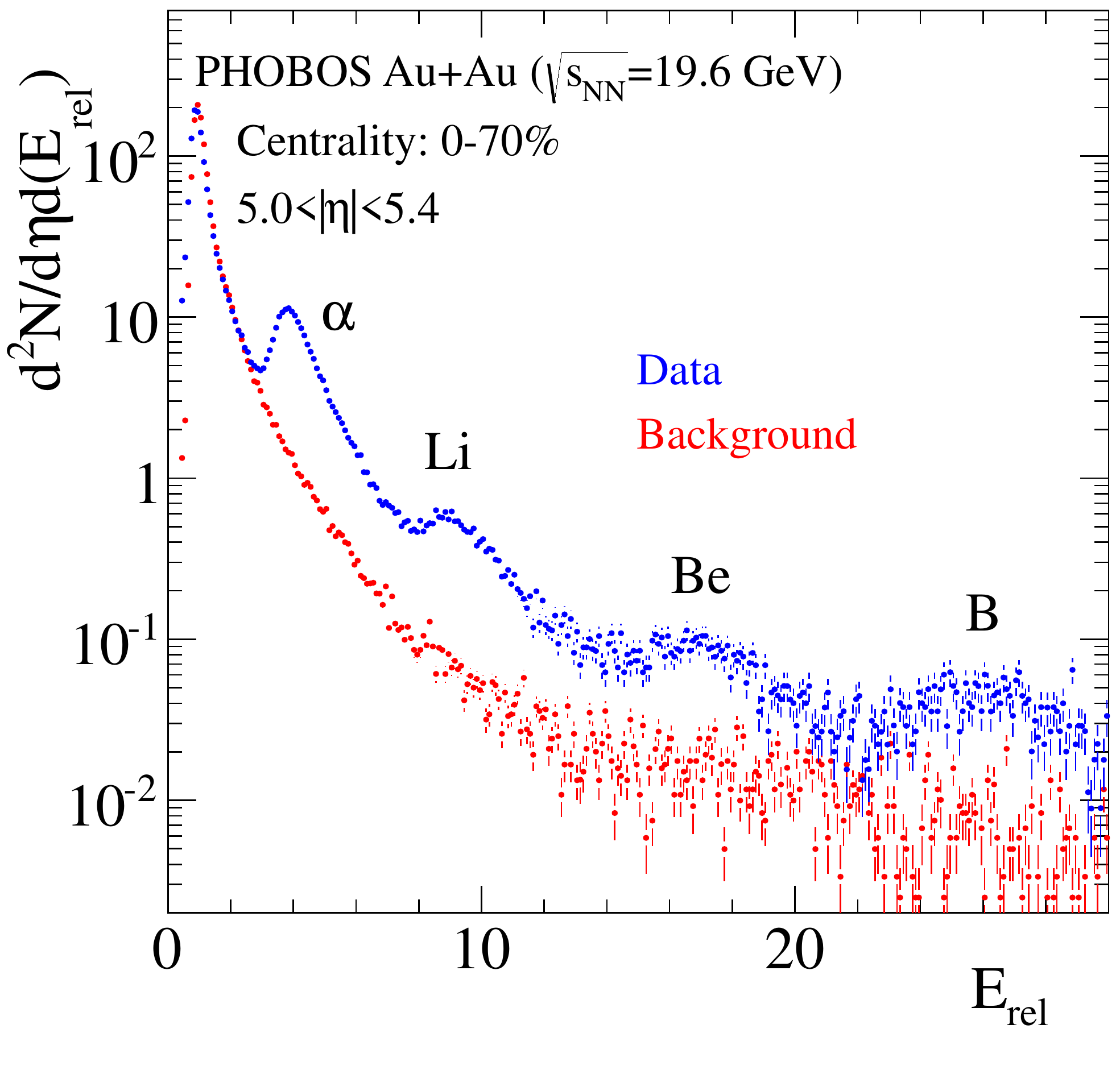}%3
 % \end{minipage}
  %\hspace{0.05\textwidth}
  %\begin{minipage}{0.34\textwidth}
    \centering
    \caption{\label{fig:AuAuMethod}
      (color online) The distribution of the relative energy loss
      in Au+Au collisions as $\snn$\,=\,19.6\,GeV averaged over the
      centrality range 0\%--70\% and 5.0$<$$|\eta$$<$5.4.
      The blue distribution shows data, the error bars indicate
      statistical uncertainties only and the data
      are not corrected for acceptance.  The red distribution shows the
      results from a MC simulation of singly-charged particles with
      spectator fragments explicitly excluded.  See text for discussion.}
  %\end{minipage}
\end{figure}

To determine the spectral shape of the $Z$\,=\,1 contribution, the energy
loss signal for single particles is modeled using a full {\sc Geant}
Monte-Carlo (MC) simulation of the PHOBOS apparatus.
In data and simulation, it is observed that multiple $Z$\,=\,1
particles can impinge on a single silicon sensor, causing an ensemble
distribution over many events to exhibit peaks at \Erel$\sim$2~and~3
(note that these additional peaks are not clearly visible in Fig.~\ref{fig:AuAuMethod}).
The peak at \Erel$\sim$2 (which occurs at a rate of about 8\% at the highest
pseudorapidities) has to be accounted for in the $Z$\,=\,1 subtraction.
The third peak is suppressed to a rate of 0.6\% and is ignored
in this analysis.  As this rate is dependent on the charged-particle
multiplicity in each detector, this fraction varies with both
centrality and pseudorapidity, an effect observed in both data and simulation.
Importantly, data with a lesser contribution from
a second charged-particle effectively steepens the spectrum, changing the
amount of subtracted background.

To account for the second peak in the spectrum, both data and MC are
divided into five pseudorapidity and seven (six) centrality classes
for the Au+Au (Cu+Cu) analysis, respectively.
As the MC distribution only reflects the relative contribution of
1 and 2 singly charged-particles, each class produces a spectrum which
has a unique shape.  To account for the contribution of a second singly charged particle,
each data class is systematically compared to all
centrality/pseudorapidity classes from the MC, i.e. 35 comparisons, therefore
testing the data against a large sample of simulated 2/1 hits-per-sensor
contributions.  Each MC class is normalized to the data at the first
peak (close to \Erel=1 in Fig.~\ref{fig:AuAuMethod}).
The optimal background is chosen as the one with the least $\chi^{2}$ difference
between data and MC \Erel spectra, formed over a region around
the expected second peak position (1.5$<$\Erel$<$2.5).

To systematically test the sensitivity of the one-to-two hits
contribution, $Z$\,=\,1 MC simulation samples with different one-to-two hits
ratios are used in the analysis.  A systematic uncertainty
due to the $\chi^{2}$ procedure is assigned by considering two
further $Z$\,=\,1 distributions.  First, the distribution
with the next-smallest $\chi^{2}$ was used, and a full
reanalysis was made.  Second, a $Z$\,=\,1 distribution with
$\chi^{2}/d.o.f.$\,=\,$\chi^{2}_{min}/d.o.f.$\,+\,1 was selected, with a
full reanalysis performed.  A systematic difference of 3\%--12\%
was found for the $Z$\,=\,2 fragment yield in Au+Au collisions in the
highest pseudorapidity bins.
In pseudorapidity and centrality bins where there is a negligible
higher-$Z$ yield, the MC class determined from this analysis
closely replicates the entire tail of the singly-charged particles.

\begin{figure}[!ht]
%  \begin{minipage}{0.55\textwidth}
    \centering
    \includegraphics[width=0.99\linewidth]{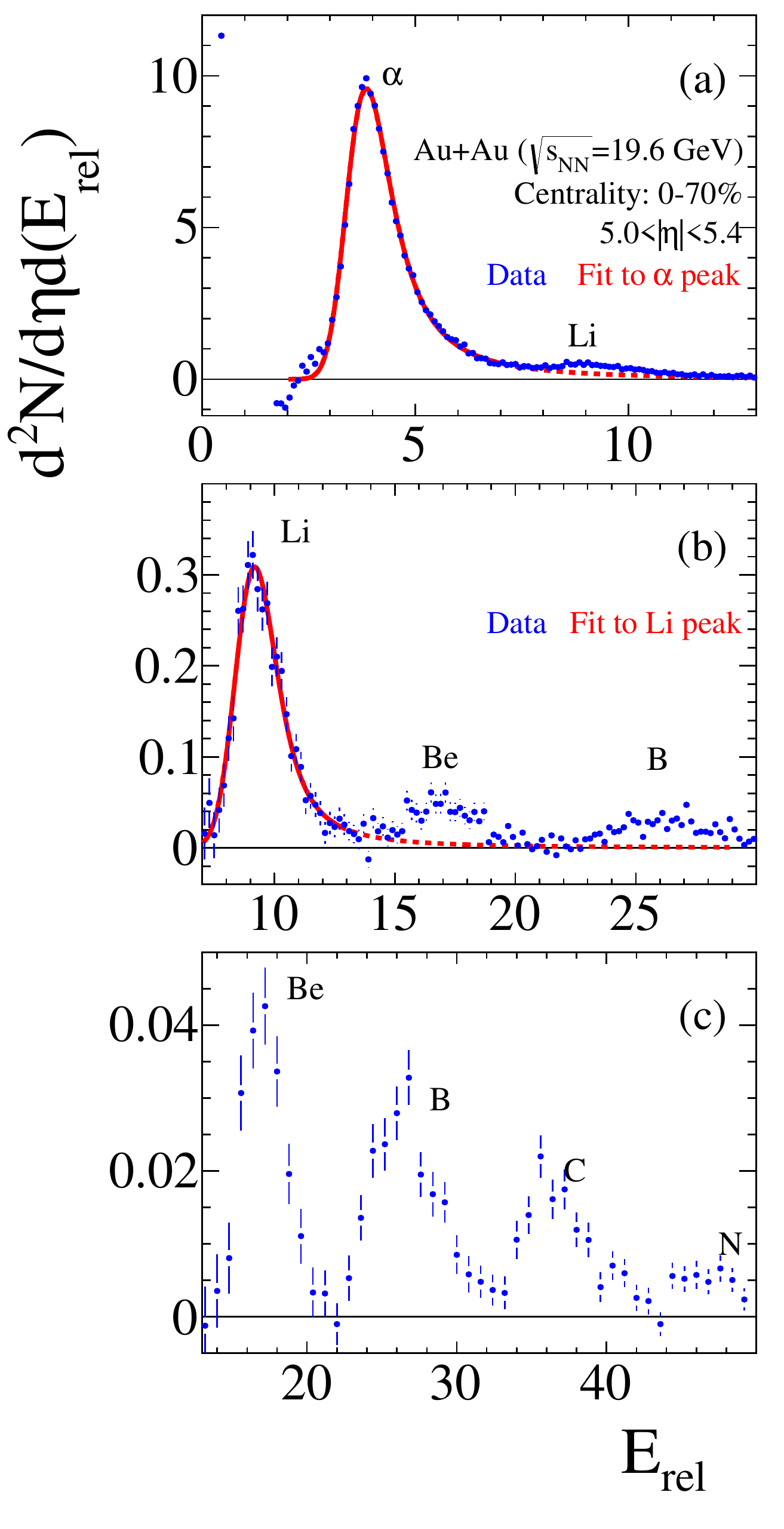}%3
 % \end{minipage}
  %\hspace{0.05\textwidth}
  %\begin{minipage}{0.34\textwidth}
    \centering
    \caption{\label{fig:AuAuMethod2}
      (color online) Panel (a) shows the \Erel distribution after subtracting the $Z$\,=\,1 component. The dominant
      peak at \Erel$\sim$4 corresponds to $Z$\,=\,2 ($\alpha$) fragments.  The red line
      depicts the fit to determine fragment yields -- the solid part shows 
      the region over which the fit was made and the dashed is the
      extrapolation under the higher-$Z$ peaks.
      Panel (b) shows the same as (a) but with the contribution from the
      $\alpha$ spectrum (red line in (a)) removed, highlighting the
      distribution from $Z$\,$\ge$\,3 fragments.  The red line shows a
      fit to the Lithium peak, similar to that described in (a).
      Panel (c) shows the same as (b), but with the contribution from
      $Z$\,=\,3 particles removed, and the $x$-axis is extended to 
      show the presence of $Z$\,=\,6 and $Z$\,=\,7 fragments. 
      The error bars are statistical only; data
      are not corrected for acceptance. See text for discussion.}
  %\end{minipage}
\end{figure}

\begin{figure*}[!t]
\centering
\includegraphics[angle=0,width=0.99\textwidth]{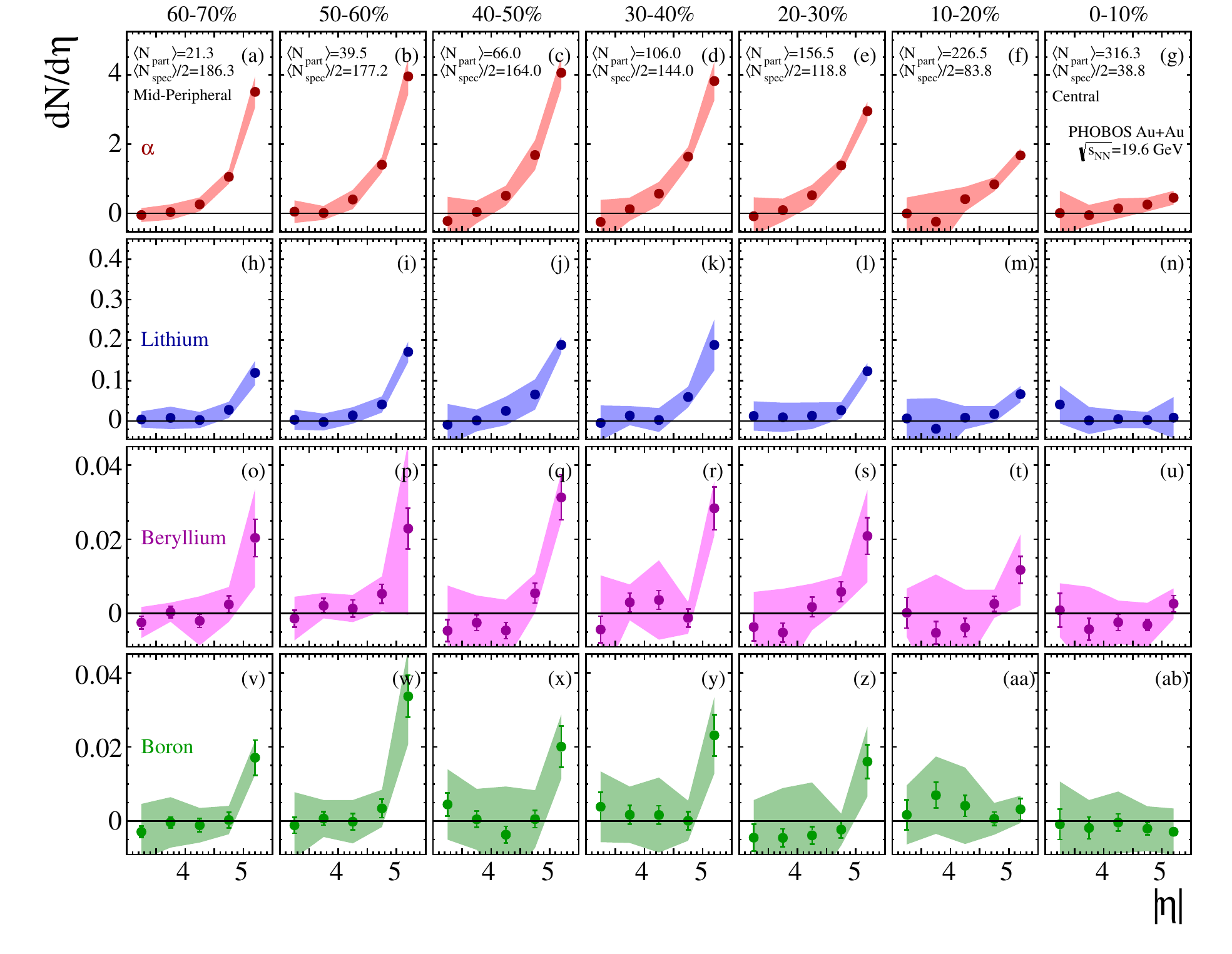}%4
\caption{\label{fig:AuAuEta}
(color online) Pseudorapidity dependence of $\alpha$ (panels
(a)-(g)), Lithium (h-n), Beryllium (o-u), and Boron
(v-ab) fragments measured in Au+Au collisions at
$\snn$\,=\,19.6\,GeV.  Data are presented in bins of
centrality (more central in the rightmost panels) and
are averaged over both hemispheres, i.e. the number of
fragments per colliding nucleus.
The error bars represent the statistical uncertainty, the
error bands represent 90\% C.L. systematic uncertainties in the yield.}
\end{figure*}

\subsection{Extracting Fragment Yields}

The measured \Erel distribution after subtraction of the fitted
$Z$\,=\,1 contribution is shown in Fig.~\ref{fig:AuAuMethod2}a.
The spectrum is 
dominated by the $Z$\,=\,2 (referred to here as $\alpha$)~\footnote{Note: $Z$\,=\,2 could imply either $^{3}$He or $^{4}$He ($\alpha$).  However, as the abundance of $^{4}$He is far greater, we refer to $Z$\,=\,2 as $\alpha$.}
fragments.  To determine the
yield, the peak is fit with a convoluted
Landau and Gaussian function (solid red line) in a region close to the
$\alpha$ peak, such that the fit range does not overlap
the region where the Lithium peak is expected.
The mean position in the fit is constrained to be the
expected mean position for the $\alpha$ fragments.
The use of a Landau function
is necessary to account for the high tail which partially resides
underneath the higher mass peaks -- in much the same way
that the tail of the singly charged particles contributed to the
$\alpha$ peak, before subtraction. The total yield is
calculated as the integral of this fit, extrapolated to
encompass $\alpha$ fragments appearing at high \Erel, for example under the
Lithium peak (shown by the dashed red line). This extrapolation ultimately contributes less
than 10\% of the total yield, and the agreement between
the raw data and the fit integrated over the same region (3\,$<$\,\Erel\,$<$\,6)
is better than 3\%.

The full $\alpha$ contribution to the energy loss spectrum is
then subtracted (red line in Fig.~\ref{fig:AuAuMethod2}a) to leave only $Z$\,$\ge$\,3 fragments
(Fig.~\ref{fig:AuAuMethod2}b). Next, with a similar
procedure, the yield of Lithium fragments
is determined using a Landau+Gaussian form (red solid and dashed lines), which is then
subtracted from the relative energy loss spectrum. For the final
distribution, $Z$\,$\ge$\,4 shown in Fig.~\ref{fig:AuAuMethod2}c,
the effect of the Landau tail is overpowered by the
Gaussian width, and thus a two-Gaussian fit is used
to extract the yields for Beryllium and Boron fragments.
The mean positions used in this fit are constrained to be
the expected position for each fragment.  The number of these
$Z$\,$>$\,3 fragments is only 1\% of $\alpha$ particles.
As such, a small constant offset is allowed to account for possible
uncertainties in subtracting $\alpha$ and Lithium contributions
to the spectrum, which could lead
to over- or under-subtraction on the spectrum.
For charges greater than five, the
full centrality and $\eta$ dependence is limited by
the statistics collected in the single day of Au+Au running at
the RHIC injection energy of $\snn$\,=\,19.6\,GeV, and are
therefore not included in this analysis. The same procedure
is used to obtain $Z$\,=\,2 and $Z$\,=\,3 fragment yields in
Cu+Cu collisions at $\snn$\,=\,22.4\,GeV;  $Z$\,$>$\,3 fragments
are not observed, even given the larger statistics of the sample.

\subsection{Corrections and Systematic Uncertainty}
The data are corrected for acceptance via simulation which compares
the number of tracks which impinge the detectors to all 
tracks in the full solid angle.  As
the $Z$\,=\,1 ``background'' is explicitly subtracted, no further
corrections are applied. The effect of absorption of
the fragments in the 1\,mm thick Beryllium
beam pipe was evaluated via a {\sc Geant} simulation and
was found to be negligible ($<$1\%) as the fragments
are high energy -- $E_{\rm fragment}$\,$\approx$\,9.8\,GeV
(11.2\,GeV) per nucleon for Au+Au (Cu+Cu) collisions.

Systematic uncertainties (90\% C.L.) are evaluated by performing
several checks, in addition to those due to the Landau
$Z$\,=\,1 background subtraction.  The difference in the extracted yields
measured independently in the positive and negative pseudorapidity
regions of the PHOBOS detector is found to be 3\%--11\% for the $\alpha$
yields in Au+Au collisions at the highest pseudorapidities, dependent
on centrality.  A shift of the 
measured energy scale in the \Erel calculation was applied ($\pm$5\%)
which results in a 1\%--8\% uncertainty on the $\alpha$ yield for the
highest pseudorapidities.  A total systematic uncertainty of 11\%
is assigned on the $\alpha$ yield for the highest pseudorapidities
in the 40\%--50\% centrality class.
For larger fragments, an additional uncertainty due to the 
subtraction of the measured $\alpha$ yield is estimated to be 1.5\% for
Lithium for the highest pseudorapidities in Au+Au collisions.
The systematic uncertainties for 40\%--50\% Au+Au collisions at
the highest pseudorapidities are 11\%, 20\%, and 45\%
for Lithium, Beryllium, and Boron, respectively.

It was also checked whether fragments could be due to
interactions between collision products and the beam pipe,
by measuring the number of $Z$\,=\,2 fragments in
$\snn$\,=\,62.4\,GeV~and~200\,GeV data.  Few were
observed in the former, while none were
observed at the highest energy.  Should the high-$Z$
fragments have emanated from dead and active detector material, notably
the Beryllium beam pipe, then 
the most central $\snn$\,=\,200\,GeV data, which has a larger multiplicity,
would have included more background than the lower energy data. Instead,
we find no evidence of $Z$\,=\,2 (or higher) fragments in the highest
energy data, indicating that such backgrounds from dead material are
negligible.

\section{Results I -- Pseudorapidity Dependence}\label{sec:EtaDep}

\begin{figure*}[!ht]
  \begin{minipage}{0.65\textwidth}
    \centering
    \includegraphics[angle=0,width=0.99\textwidth]{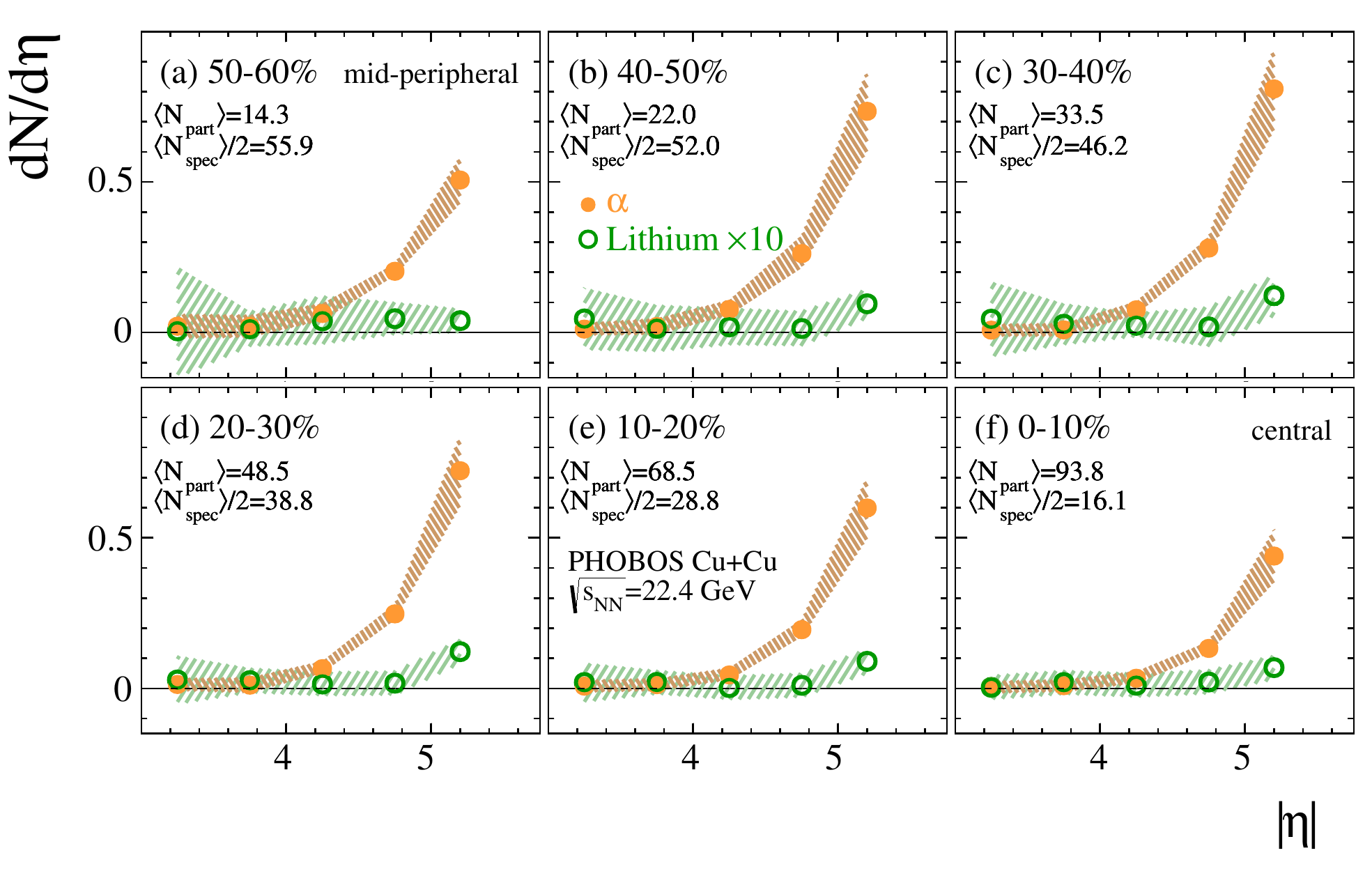}%5
  \end{minipage}
  \hspace{0.05\textwidth}
  \begin{minipage}{0.24\textwidth}
    \centering
    \caption{\label{fig:CuCuEta}
      (color online) Pseudorapidity dependence of $\alpha$ (filled
      symbols) and Lithium fragments (open symbols) measured in Cu+Cu
      collisions at $\snn$\,=\,22.4\,GeV.  Lithium fragment yields
      are scaled up by a factor of 10 for clarity.  Data are presented in
      bins of centrality and are averaged over both
      hemispheres, i.e. the number of fragments per colliding nucleus.
      The error bars represent the statistical uncertainty, the
      error bands represent 90\% C.L. systematic uncertainties in the yield.\\\\}
  \end{minipage}
\end{figure*}

Both the Au+Au and Cu+Cu data are divided into five bins of
pseudorapidity and seven and six bins of centrality, respectively,
corresponding to the top 70\% (60\%) of nuclear inelastic
cross-section.  Figure~\ref{fig:AuAuEta}
shows the measured fragment multiplicity, $dN/d\eta$, as a function of pseudorapidity
(tabulated data are included in Appendix~\ref{ap:Tables}), averaged
over both hemispheres (i.e. the number of fragments per
colliding nucleus) for Au+Au collisions at $\snn$\,=\,200\,GeV.  The first row corresponds
to $\alpha$ fragments. Li, Be, and B fragments are
shown in subsequent rows. The most central data (those with
the least number of spectators after the collision) are shown in the
rightmost column; the most
peripheral are shown in the leftmost column. As is
apparent from this figure, there are no $Z$\,$>$\,1
fragments for low pseudorapidities ($|\eta|$$<$4.0) and only a
small number of fragments are produced at high centrality (0\%--10\% central).
The lightest fragment measured ($\alpha$) is observed
in each of the last three $|\eta|$ bins, Lithium fragments
are observed in the highest two bins, and Beryllium and Boron
fragments are seen only in the highest $|\eta|$ bin.

Figure~\ref{fig:CuCuEta} shows the measured $dN/d\eta$ for 
$\alpha$ and Lithium fragments in Cu+Cu collisions
at $\snn$\,=\,22.4\,GeV -- note that Lithium
yields are scaled up by a factor of 10 for clarity.  Similarly
to the Au+Au results, no spectator fragments are observed in
the low pseudorapidity region; Lithium fragments are only
observed in the highest pseudorapidity bins.

\subsection{Comparison to Charged-particle pseudorapidity density}

\begin{figure}[!t]
\centering
\includegraphics[angle=0,width=0.445\textwidth]{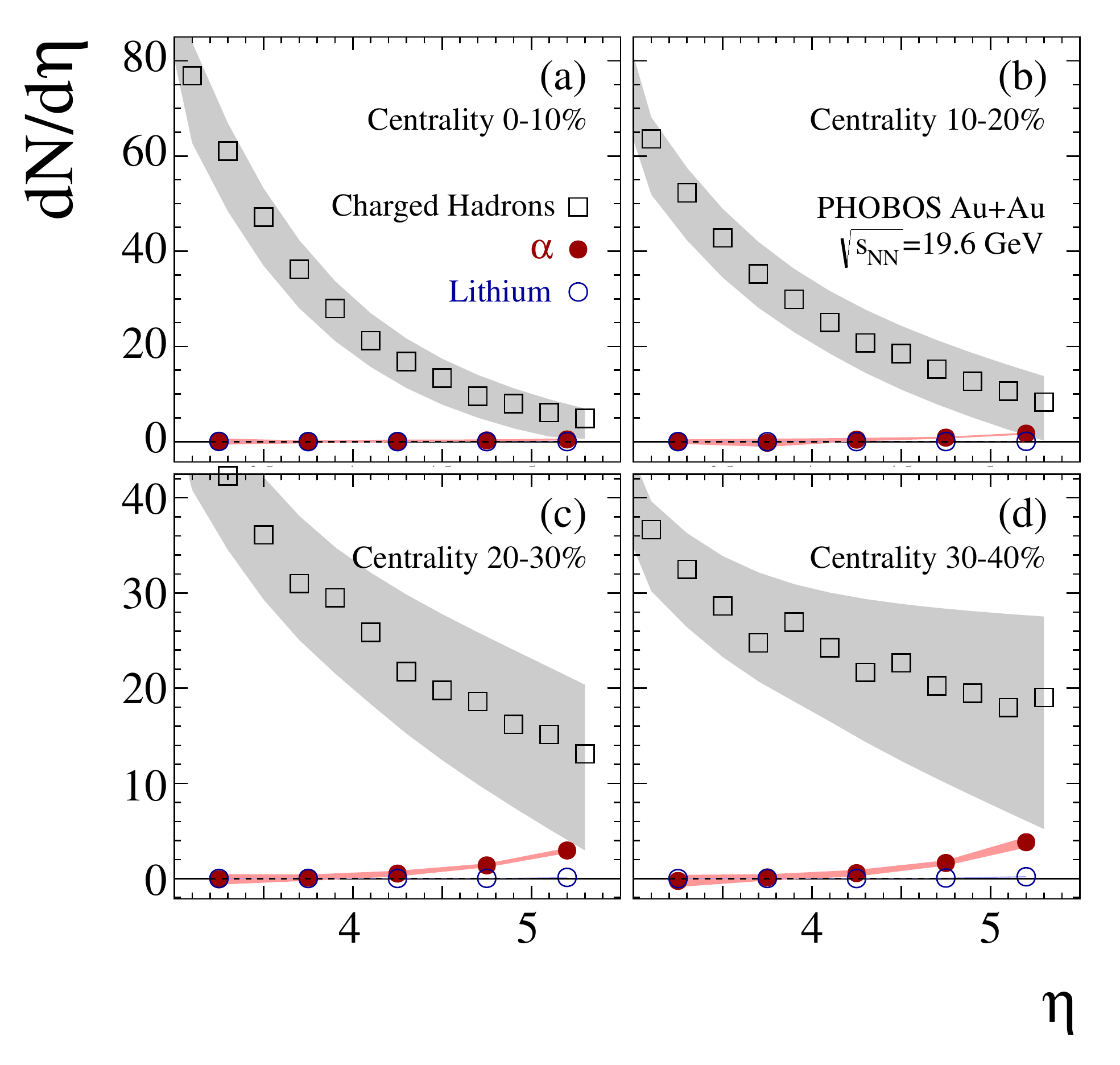}%6
\caption{\label{fig:CompChHadAuAu}
(color online) Comparison between the PHOBOS charged particle
multiplicity measured at positive $\eta$ in Au+Au collisions at
$\snn$\,=\,19.6\,GeV
and the yield of $\alpha$ and Lithium fragments, averaged over positive and negative $|\eta|$.
Panels (a), (b), (c), and (d) show the distributions in
centrality bins 0\%--10\%, 10\%--20\%, 20\%--30\%, and 30\%--40\%,
respectively.  The open squares/light grey bands
represents the PHOBOS
multiplicity~\cite{cite:BigMultPaper}, filled (open)
circles represent the measured $\alpha$ (Li) yields.}
\end{figure}

\begin{figure}[!t]
\centering
\includegraphics[angle=0,width=0.445\textwidth]{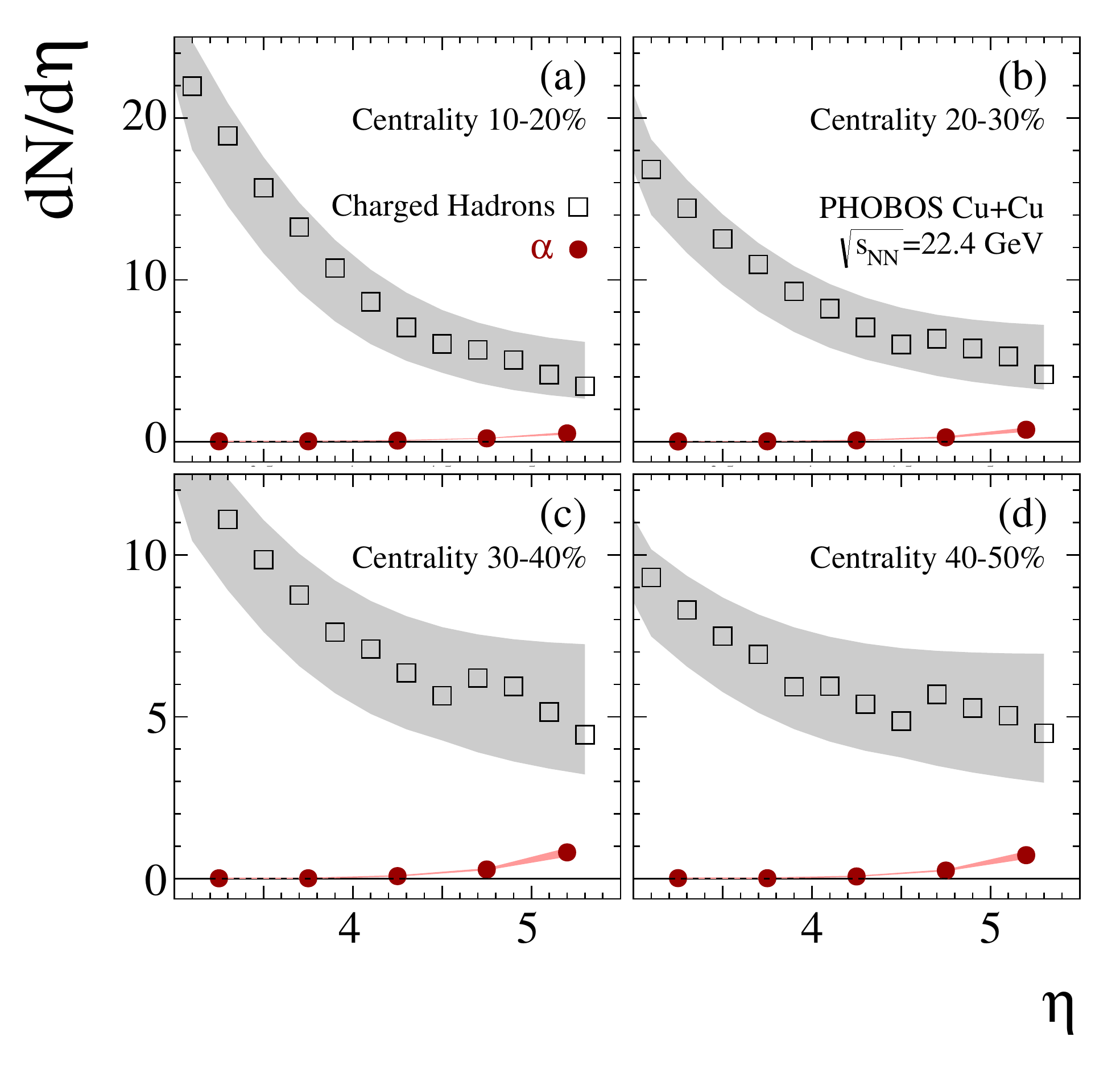}%7
\caption{\label{fig:CompChHadCuCu}
(color online) Comparison between the measured PHOBOS charged particle
multiplicity in Cu+Cu collisions at $\snn$\,=\,22.4\,GeV
and the yield of $\alpha$ fragments.
Panels (a), (b), (c), and (d) show the distributions in
centrality bins 10\%--20\%, 20\%--30\%, 30\%--40\%, and 40\%--50\%,
respectively.  The open squares/light grey bands
represents the PHOBOS multiplicity~\cite{cite:BigMultPaper},
filled circles represent the measured $\alpha$ yields.}
\end{figure}

PHOBOS has measured charged particle production in the very forward region
($|$$\eta$$|$$>$$\sim$3) for Au+Au and Cu+Cu collisions~\cite{cite:SigFrag,cite:PHOBOS_AuAuCuCuMult,cite:BigMultPaper}.
It was observed that the yield of charged particles in this forward
pseudorapidity region is larger
in the most peripheral collisions compared to the central ones.
In those analyses, no distinction was made between singly- and
multiply-charged particles, so it was unclear how many of these particles were
protons (or deuterons or tritons) and how many were multiply-charged fragments. 
Figure~\ref{fig:CompChHadAuAu}
(\ref{fig:CompChHadCuCu}) shows a comparison between the
pseudorapidity-averaged $\alpha$ yield in Au+Au (Cu+Cu) collisions measured in
this analysis and the charged-particle
multiplicity ($\eta$\,$>$\,3) from the prior PHOBOS
analyses~\cite{cite:BigMultPaper}.
For these centrality bins, the yield of multiply-charged spectator
fragments for both systems is typically small
(dN$_\alpha$/d$\eta$\,=\,3.8\,$\pm$\,0.6 in 30\%--40\%
central collisions at $\snn$\,=\,19.6\,GeV)
compared to the
total charged-particle multiplicity (18.5\,$^{+9.2}_{-12.5}$). 
Therefore, % with the possible
%exception of the most forward pseudorapidity bin for Au+Au,
the majority of the particles in the forward region included in the previously
published analyses are singly-charged.
Averaged over centrality, the small abundance of multiply-charged
relative to singly-charged particles at the highest pseudorapidity
is also clearly seen in Fig.~\ref{fig:AuAuMethod}.

\subsection{Comparison to Other Fragment Data}

\begin{figure}[!t]
  \centering
  \includegraphics[angle=0,width=0.435\textwidth]{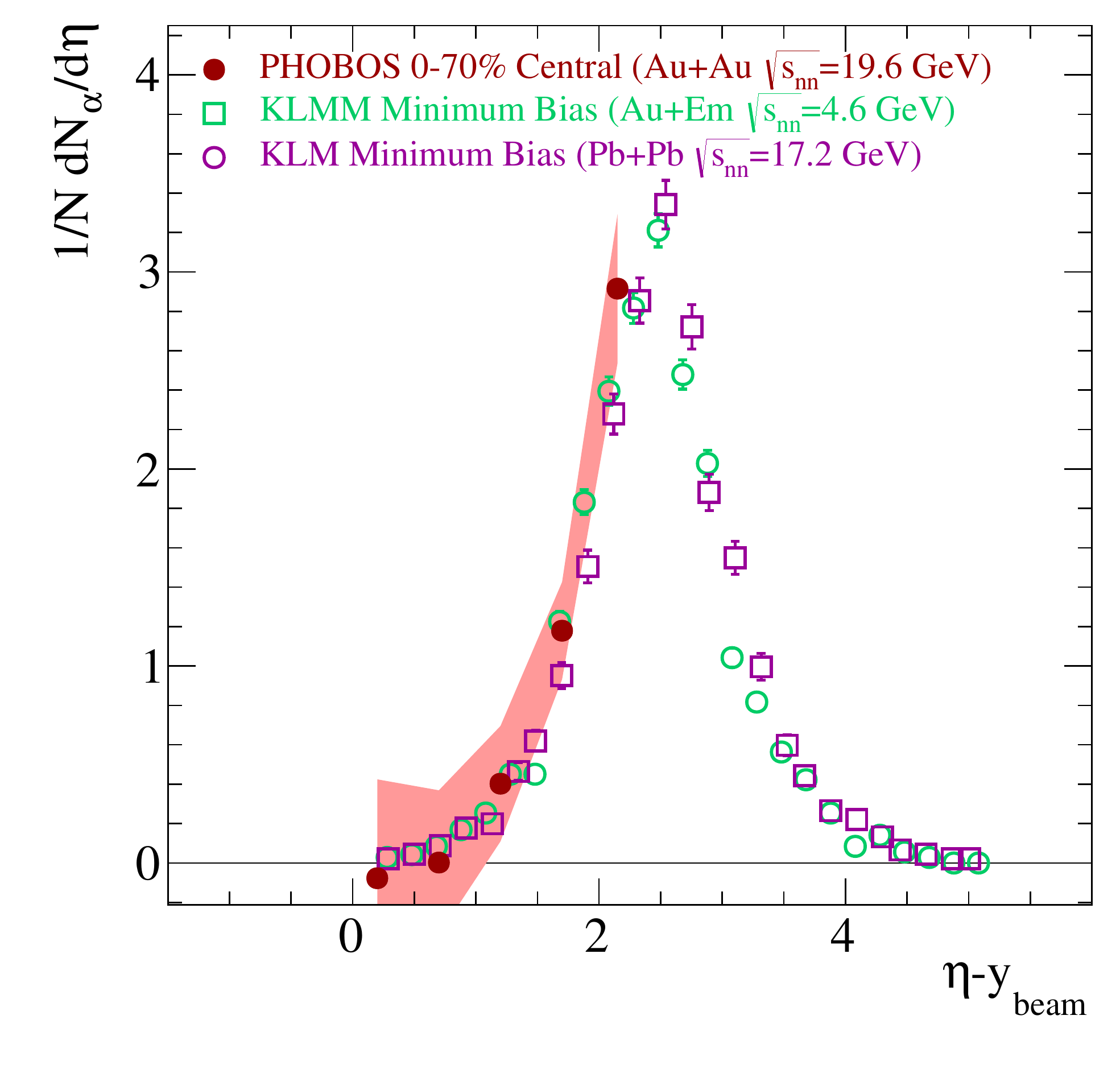}%8
  \caption{\label{fig:EnergyComparison}
    (color online) Comparison of $\alpha$ yields between PHOBOS data from Au+Au collisions
    ($\snn$\,=\,19.6\,GeV) and Au+Em ($\snn$\,=\,4.6\,GeV)~\cite{cite:KLMM_PhysRevC}  and
    Pb+Pb ($\snn$\,=\,17.2\,GeV)~\cite{cite:KLM_ActaPhysPol} collisions.
    PHOBOS data are averaged over positive and negative $\eta$ and over
    the most central 0\%--70\% cross-section
    (filled circles and shaded band which represent the 90\% C.L. systematic
    uncertainties in the yield) for $\alpha$ particles.
    The pseudorapidity ($x$-axis) is relative to the rest
    frame of the target nucleus for each energy, as discussed in Appendix~\ref{ap:EtaPrimer}.}
\end{figure}

The number of $\alpha$ particles measured by PHOBOS is
found to be similar to the yields measured in other
experiments. Figure~\ref{fig:EnergyComparison} compares
the measured $dN_{\alpha}/d\eta$ from PHOBOS (filled 
circles with a band representing the 90\% C.L. systematic
uncertainties in the yield)
with that from the KLMM~\cite{cite:KLMM_PhysRevC}
(Au projectile with beam energy 10.6\,GeV per nucleon on a
fixed emulsion (Em) target) and KLM~\cite{cite:KLM_ActaPhysPol}
(Pb projectile with beam energy 158\,GeV per nucleon on a fixed
Pb target) collaborations\footnote{The error bars shown for KLM and KLMM
data in
Fig.~\ref{fig:EnergyComparison} are based on the number
of counts, $N$, in each $\eta$ bin as $\sqrt{N}$.}.
Note that the PHOBOS
data are effectively a collision of a Au projectile with
$E_{\rm beam}$\,=\,9.8\,GeV per nucleon on a target Au nucleus (albeit
moving) where this energy is that of a single beam in
the collider, i.e. $\snn$/2. The data are shifted along
the $x$-axis in Fig.~\ref{fig:EnergyComparison} by the corresponding beam rapidity in each
case.
A detailed discussion of the properties of this shifted variable
($\eta'$\,=\,$\eta-y_{beam}$ or for symmetric collisions $\eta'$\,=\,$|\eta|-y_{beam}$) is given in Appendix~\ref{ap:EtaPrimer}.
Any impact of the difference of collision energy should be fully
compensated by this beam rapidity shift, however as neither the
collision systems nor the event selection are identical some
systematic differences are expected.  Small
differences in yield between Au+Au and Pb+Pb might arise
from the fact that the Pb+Pb collisions from the KLM analysis are
on average more peripheral (covering 0\%--100\%) than the Au+Au
collisions (0\%--70\%) from this analysis.  As such, any excess
yield in the PHOBOS
measurements might be due to the missing 30\% of the most
peripheral events in this data set.  Moreover, we do not see any
additional systematic effect between our data and the KLMM 
data that collided Au nuclei on Em (comprising much smaller nuclei:
H, He, C, Ag, and Br).

Although a large part of the $\alpha$ yield is outside
the acceptance of PHOBOS, the yield in the measured region
agrees reasonably well between experiments, and also illustrates the
relevance of limiting fragmentation for
spectators~\cite{cite:SigFrag}.  While Appendix~\ref{ap:EtaPrimer}
carefully describes why beam rapidity is an appropriate scale 
to shift data at different energies, it is more intuitive to 
compare boost-invariant quantities such as $dN/dp_{T}$.
Appendix~\ref{ap:dNdpt} estimates a conversion of the presented data
into $dN/dp_{T}$ as a function of $p_{T}$, and compares the resulting
distributions with those estimated from lower energy collisions, see Fig.~\ref{fig:dNdpt}.
The Cu+Cu data are not shown as the
expected difference in yield between Au (197) fragments
and Cu (63) fragments is large because of the difference in
mass -- whereas the difference
between Au (197) and Pb (208) should be negligible. 

\begin{figure*}[!t]
\centering
\includegraphics[angle=0,width=0.95\textwidth]{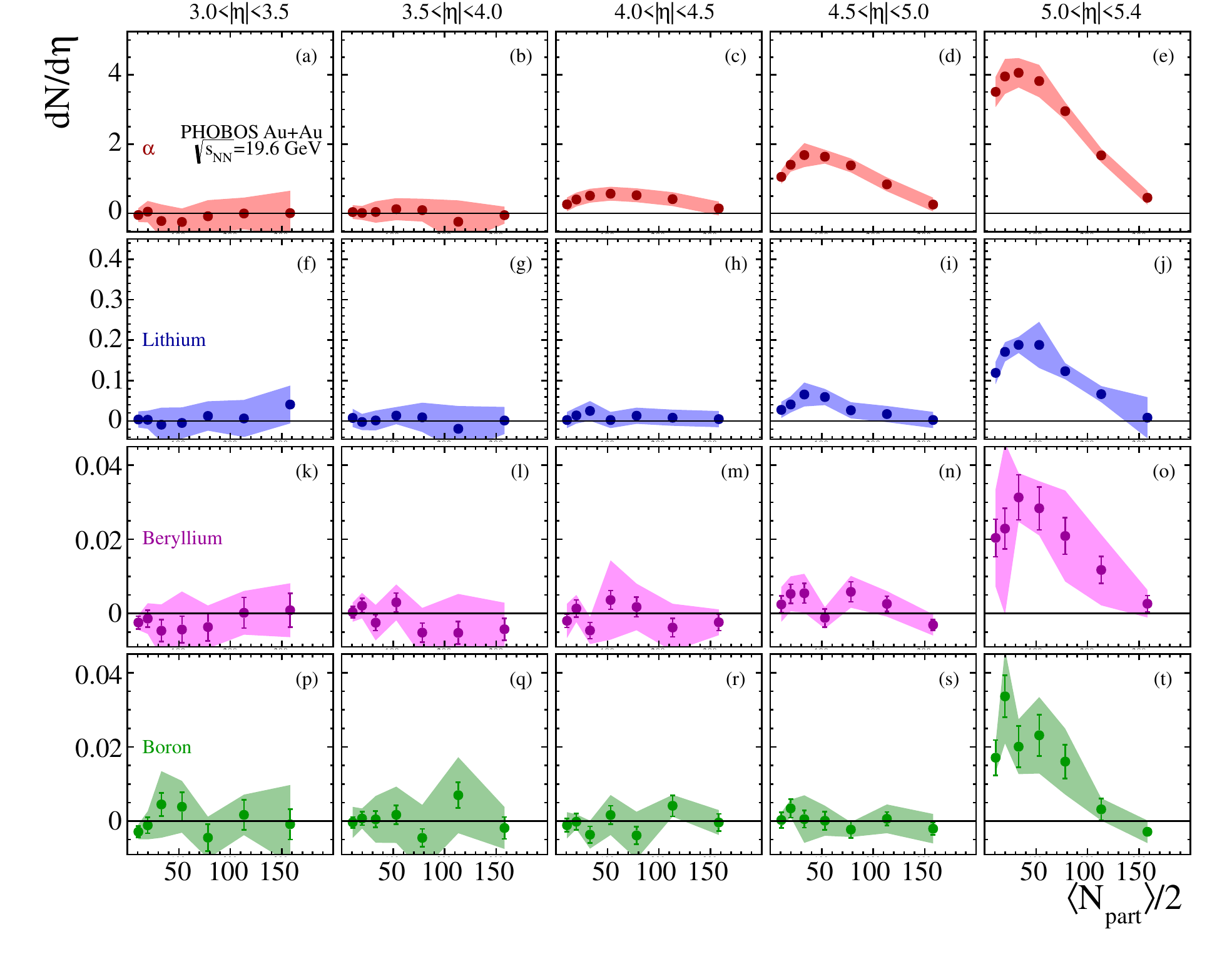}%9
\caption{\label{fig:AuAuCentrality}
(color online) Centrality dependence of $\alpha$ (panels (a)-(e)),
Lithium (f-j), Beryllium (k-o), and Boron (p-t) fragments
measured in Au+Au collisions at $\snn$\,=\,19.6\,GeV.  Data
are presented in bins of pseudorapidity, $\eta$, with the lowest
$\eta$ shown in the leftmost panels.  The data are
averaged over both hemispheres, i.e. the number of
fragments per colliding nucleus. The error bars
represent the statistical uncertainty, the error bands represent
90\% C.L. systematic uncertainties in the yield.  The 
errors associated with the centrality variables (here $\npart/2$) are not shown
on the figures, see Tables~\ref{tbl:AuAuAlphas}--\ref{tbl:CuCuLithium}. }
\end{figure*}

\section{Results II -- Centrality Dependence}\label{sec:CentDep}

Another way to look at this data is to examine the
centrality dependence, shown in
Fig.~\ref{fig:AuAuCentrality} for Au+Au collisions
at $\snn$\,=\,19.6\,GeV. Here, the absence of
fragments at low pseudorapidity is highlighted in
the first two columns.
Each $|\eta|$ bin with a significant signal (panels c-e, i-j, o, t)
shows a similar pattern:
an increase of the yield for peripheral events, a turn-over
for mid-central events, and finally an almost linear
decrease with $\npart/2$ toward the fully overlapping
collisions.  A similar dependence is also
seen in the measured ZDC energy distribution versus centrality in
the peripheral region at very high pseudorapidity, see for
example Ref.~\cite{cite:PHOBOS_WhitePaper}.

\begin{figure}[!t]
\centering
\includegraphics[angle=0,width=0.475\textwidth]{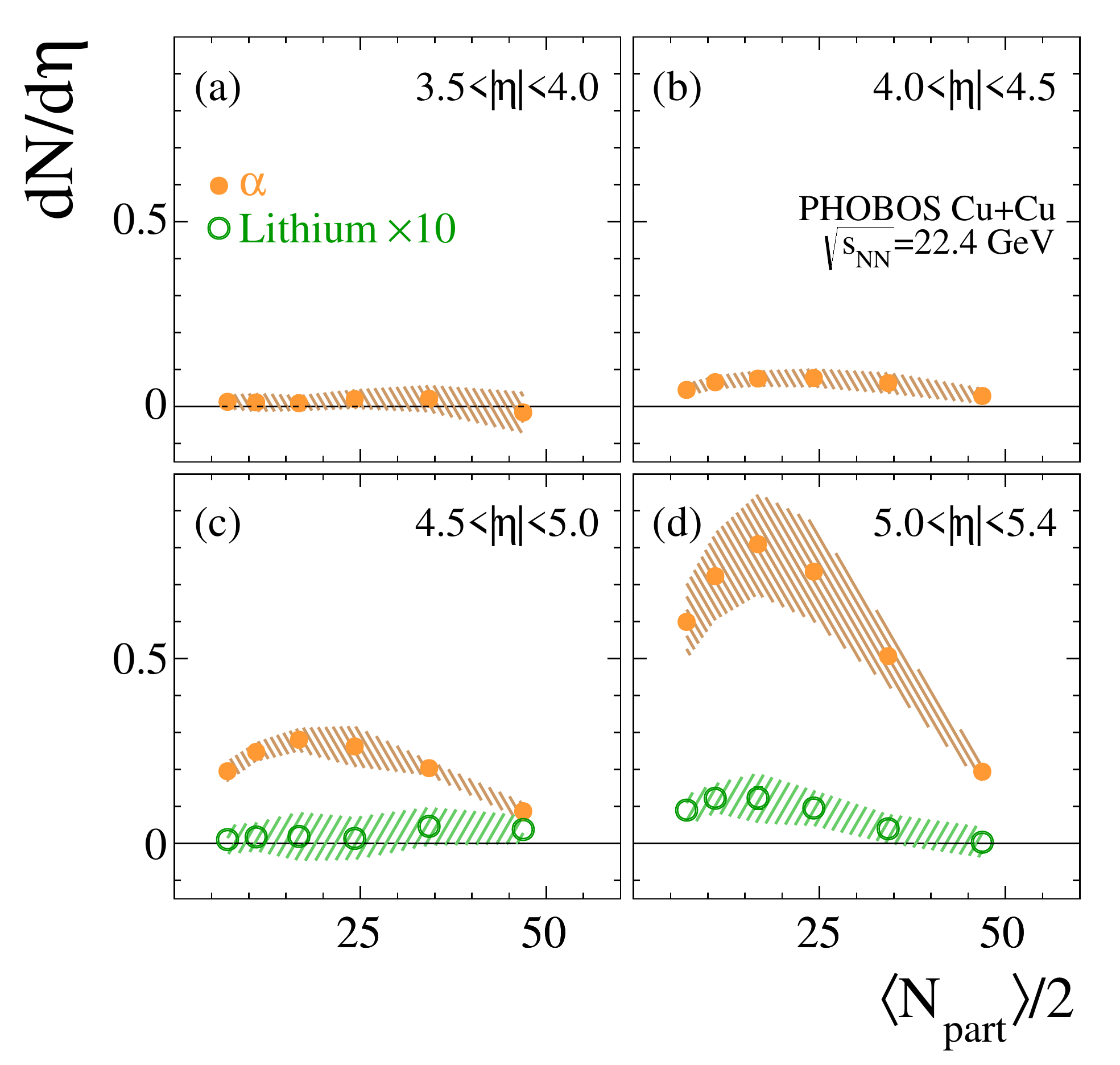}%10
\caption{\label{fig:CuCuCentrality}
(color online) Centrality dependence of $\alpha$-fragments (filled symbols) for
$\snn$\,=\,22.4\,GeV Cu+Cu collisions for four $|\eta|$
bins (a-d).  For clarity, Lithium (open symbols) are scaled up by a factor
of~10 and are only shown for the highest two pseudorapidity bins
(panels (c) and (d)).
The data are averaged over both
hemispheres, i.e. the number of fragments per colliding
nucleus.  The error bars (typically smaller than the symbol height)
represent the
statistical uncertainty, the error bands represent
90\% C.L. systematic uncertainties in the yield.}
\end{figure}

In Cu+Cu collisions at $\snn$\,=\,22.4\,GeV, a similar
centrality dependence is observed for $\alpha$
and Lithium fragments in Fig.~\ref{fig:CuCuCentrality}.
%The statistical fluctuations
%prohibit the extraction of the $Z$\,$\ge$\,4
%fragments.  
%Beryllium and Boron fragments were seen in the minimum bias
%sample, but their observed yields are too small to study
%the centrality and pseudorapidity dependencies.

\subsection{Comparison of Au+Au and Cu+Cu data}

It should be noted that the
relative coverage ($\eta'\,\equiv\,|\eta|$\,--\,$y_{\rm beam}$)
of the detector is not quite the same for Au+Au and Cu+Cu
collisions owing to the different beam
rapidities: $y_{\rm beam}$\,=\,3.04 (3.18) for Au+Au (Cu+Cu).
Therefore, in comparing the two data sets, data points are evaluated
at the same average $\eta'$,
via an interpolation between measured points.

\begin{figure}[!ht]
\centering
\includegraphics[angle=0,width=0.445\textwidth]{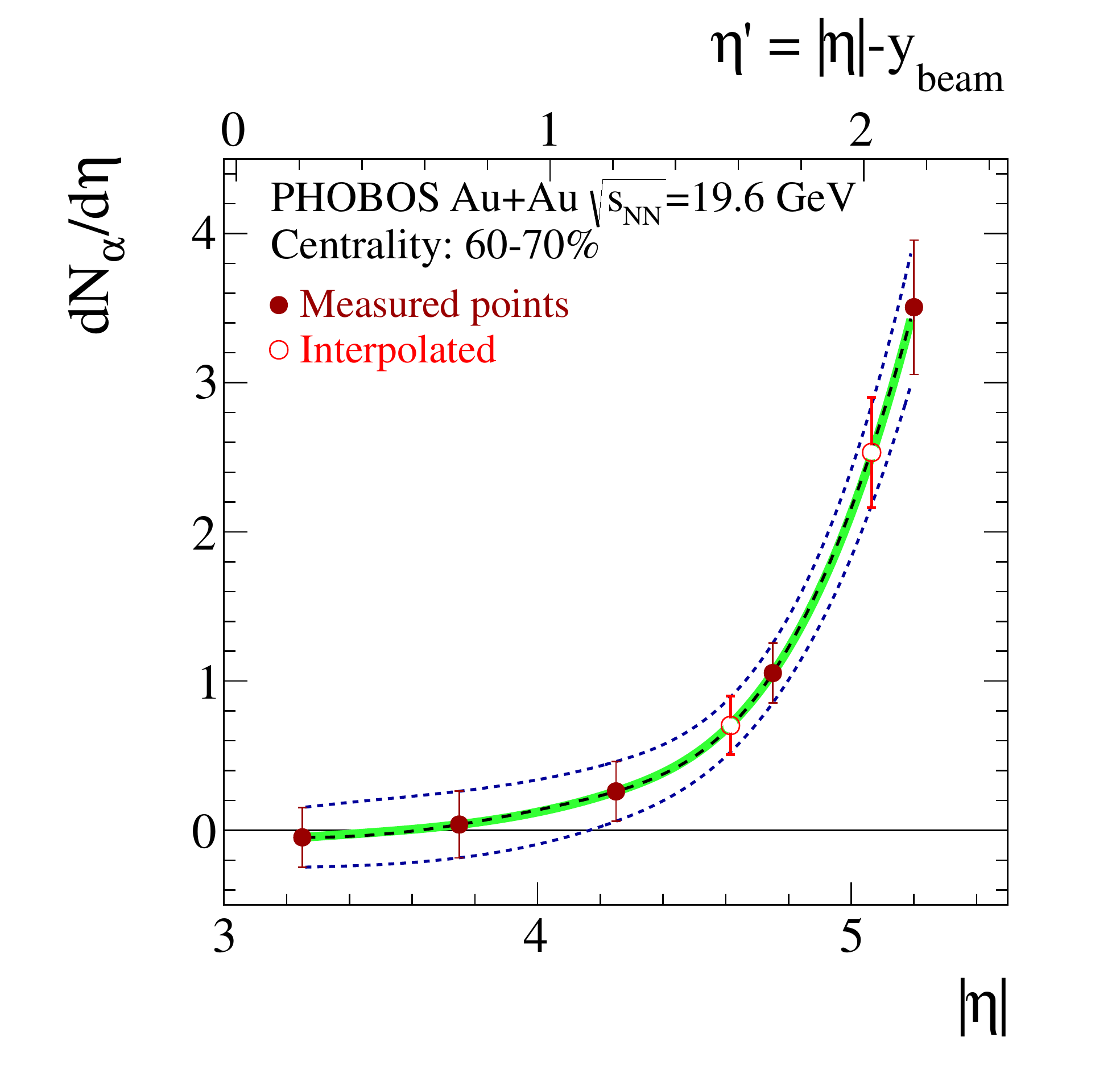}%11
\caption{\label{fig:SplineFit}
  (color online) Spline polynomial fits (lines) to $\alpha$ yields from Au+Au peripheral (60\%--70\%) data (filled
  circles).  Interpolated points at $\eta'$\,=\,1.57 and $\eta'$\,=\,2.02
  are shown as open circles. The scale on
  the upper $x$-axis shows $\eta'$\,$\equiv$\,$|\eta|$\,--\,$y_{\rm beam}$.
  The dashed and green lines show fits using polynomials of different order.
  The outer dotted lines represent a fit to points at the extreme
  of the systematic uncertainty bands.}
\end{figure}

To evaluate the yield at each $\eta'$, a polynomial spline fit is
made which smoothly connects the measured data points.
The uncertainty in this method is evaluated with two different
fits, which are found to be within 10\% of the associated data point
systematic uncertainty.  Figure~\ref{fig:SplineFit} shows an example of a fit 
to peripheral (60\%--70\%) Au+Au ($dN_{\alpha}/d\eta$) data to determine interpolated
points at $\eta'$\,=\,1.57 and $\eta'$\,=\,2.02.  A similar fit is
made to Cu+Cu data to determine an interpolated point at
$\eta'$\,=\,1.21.

\begin{figure}[!ht]
\centering
\includegraphics[angle=0,width=0.445\textwidth]{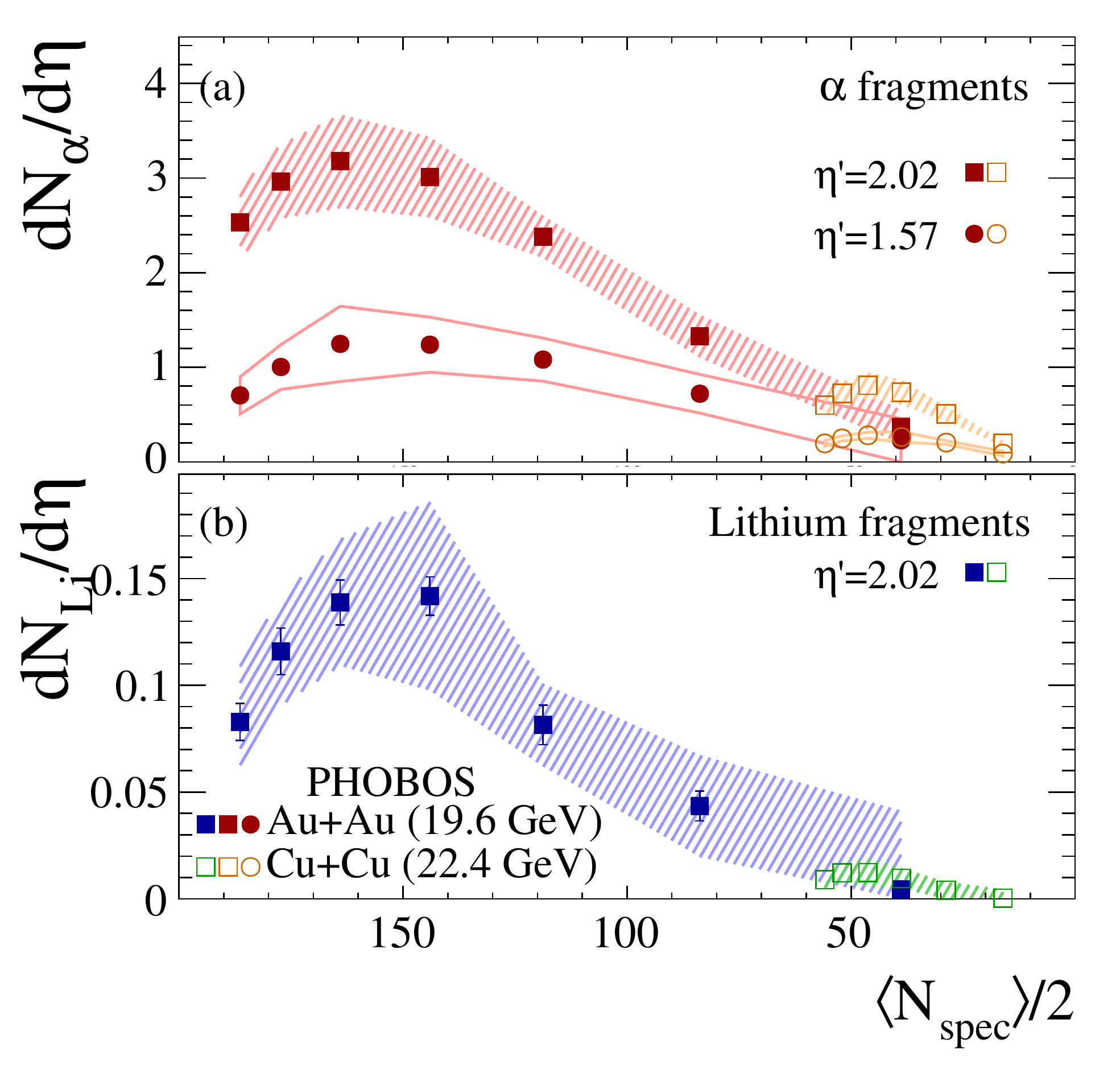}%11
\caption{\label{fig:NSpec}
(color online) Centrality dependence of $\alpha$ (panel (a))
and Lithium yields (b) in $\snn$\,=\,19.6\,GeV Au+Au
(filled symbols) and 22.4\,GeV Cu+Cu
(open symbols) collisions.
Note that the centrality variable is not $\npart/2$ but
$N_{\rm spec}$ from a
single nucleus -- see text for details -- and the $x$-axis
runs backwards, central
collisions are the rightmost data points. The $\alpha$ data
are evaluated at $\eta'$\,=\,1.57 (circles/unfilled systematic bands) and
$\eta'$\,=\,2.02 (squares/filled systematic bands). Lithium yields are only
shown for $\eta'$\,=\,2.02. The bands represent 90\% C.L.
systematic uncertainties in the yield. }
\end{figure}

A comparison of the centrality dependence of $\alpha$ and
Lithium yields for Au+Au and
Cu+Cu is given in Fig.~\ref{fig:NSpec}.
The data are averaged over both hemispheres,
representing the fragments from a single
Gold (or Copper) nucleus.  The yield of $\alpha$ and
Lithium fragments are shown versus $N_{\rm spec}$/2 from a single nucleus.
Note that the $x$-axis is inverted such that central collisions are located
rightmost on the figure.
The magnitude of the yields of fragments is proportional to
$N_{\rm spec}/2$ over a wide range of number of
spectators.
This behavior provides a simple
explanation for the smaller number of fragments observed in
peripheral Cu+Cu collisions compared to those from peripheral Au
fragmentation.
Modulo the drop-off for the most peripheral collisions,
yields are approximately similar in the two systems
for similar $N_{\rm spec}/2$.

There is some evidence that, at the same $N_{\rm spec}/2$,
the yield of $\alpha$ fragments is higher 
in Cu+Cu than in Au+Au, which is not apparent for Lithium.
This is possibly due to a preference for emitting smaller fragments
in the smaller Copper nucleus. 

\begin{figure*}[!ht]
\centering
\includegraphics[angle=0,width=0.75\textwidth]{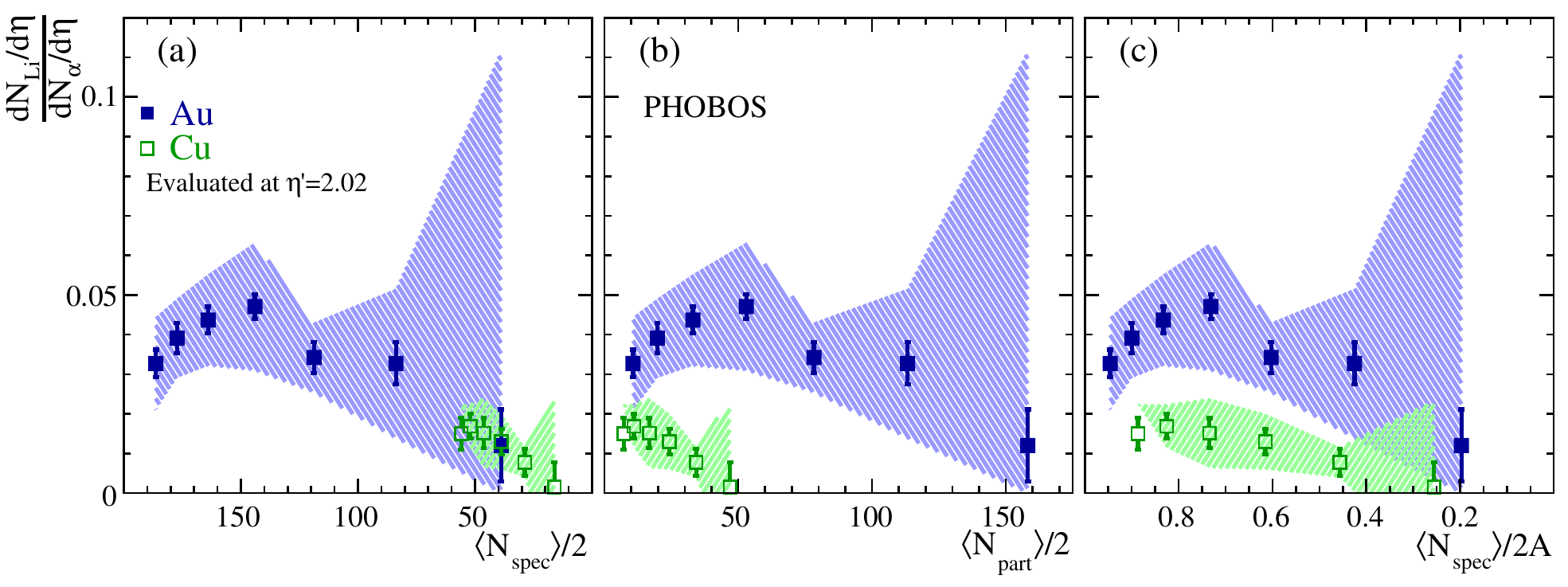}%12
\caption{\label{fig:NSpecRatioLi}
(color online) Centrality dependence of the yield of Lithium nuclei divided by that of $\alpha$ particles
evaluated at $\eta'$\,=\,2.02.
Au+Au (filled symbols) and Cu+Cu (open symbols) collision data are shown 
as a function of (a) $N_{\rm spec}/2$, (b) $N_{\rm part}/2$, and (c) the collision geometry ($N_{\rm spec}/2A$).
The bands represent 90\% C.L. systematic uncertainties in the ratio. }
\centering
\includegraphics[angle=0,width=0.75\textwidth]{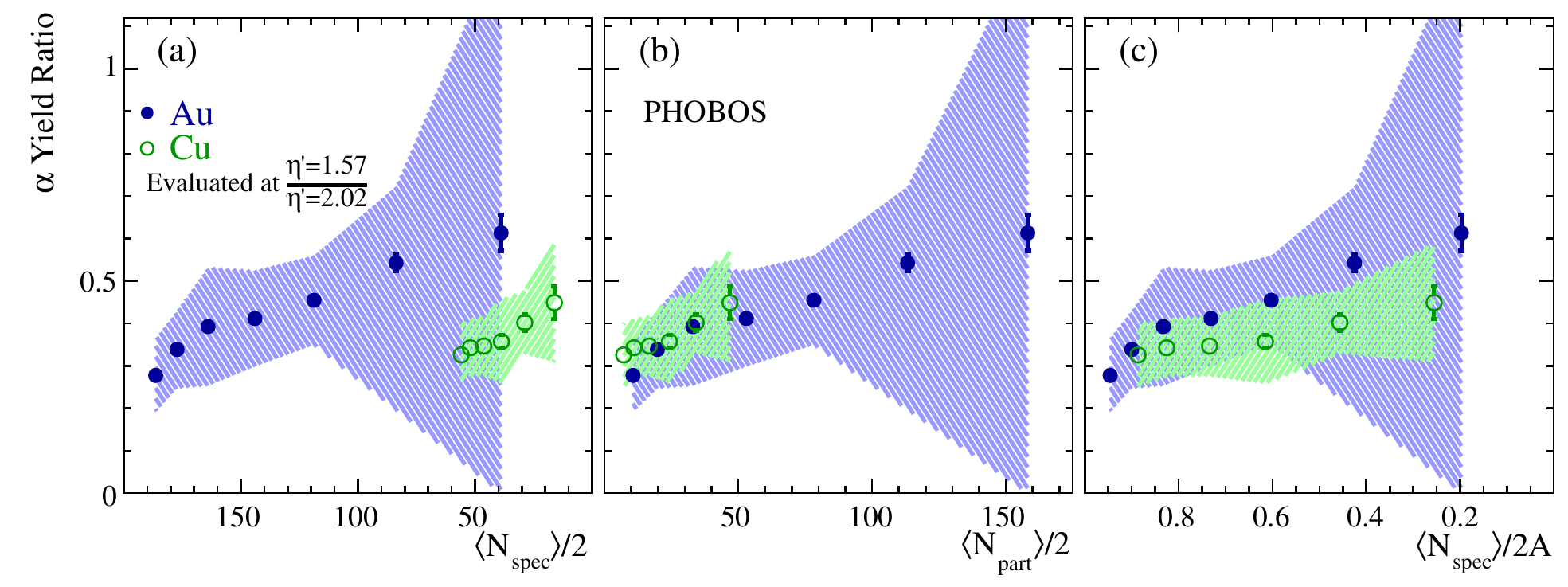}%13
\caption{\label{fig:NSpecRatio}
Centrality dependence of the yield of $\alpha$-particles evaluated at
$\eta'$\,=\,1.57 divided by the yield measured at $\eta'$\,=\,2.02.
Au+Au (filled symbols) and Cu+Cu (open symbols) collision data are shown 
as a function of (a) $N_{\rm spec}/2$, (b) $N_{\rm part}/2$, and
(c) the collision geometry ($N_{\rm spec}/2A$).
The bands represent 90\% C.L. systematic uncertainties in the ratio.}
\includegraphics[angle=0,width=0.75\textwidth]{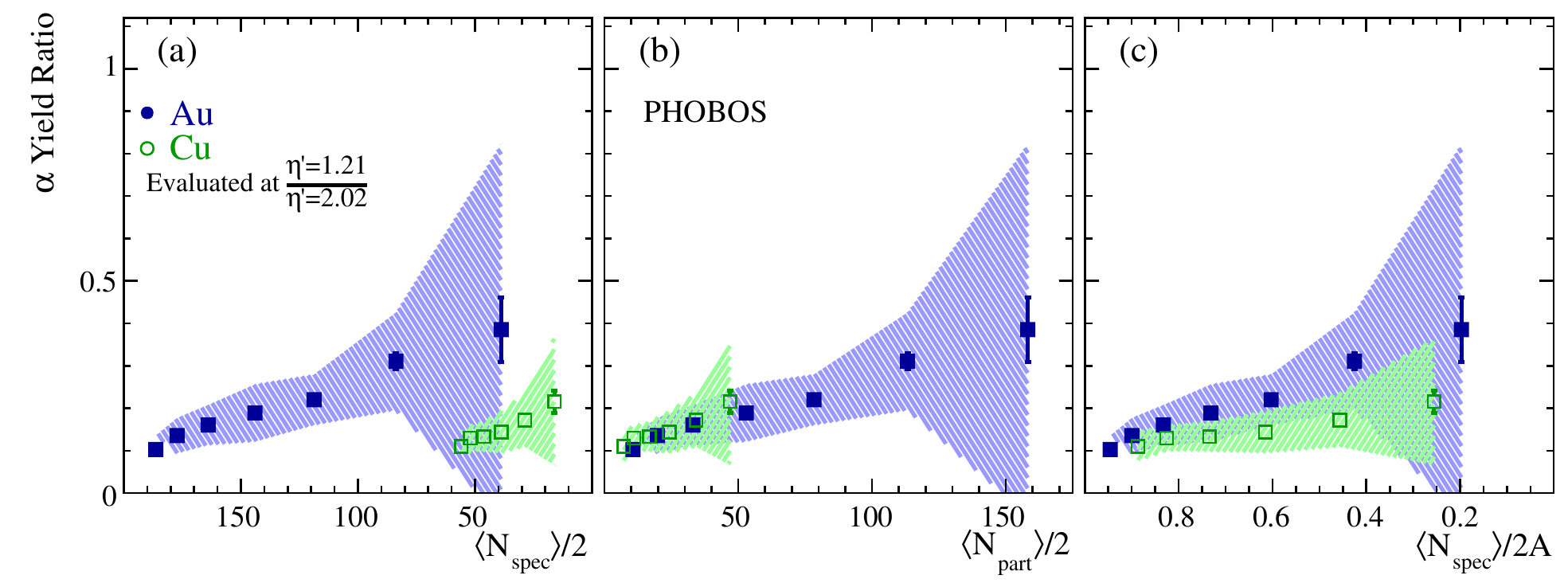}%13
\caption{\label{fig:NSpecRatio2}
Centrality dependence of the yield of $\alpha$-particles evaluated at
$\eta'$\,=\,1.21 divided by the yield measured at $\eta'$\,=\,2.02.
Au+Au (filled symbols) and Cu+Cu (open symbols) collision data are shown 
as a function of (a) $N_{\rm spec}/2$, (b) $N_{\rm part}/2$, and (c) the collision geometry ($N_{\rm spec}/2A$).
The bands represent 90\% C.L. systematic uncertainties in the ratio.}
\end{figure*}

\subsection{Pseudorapidity and Centrality Dependence of Yields}\label{sec:EtaCent}

The simultaneous pseudorapidity and centrality dependencies
of the yields can be explored by use of ratios of data, to 
investigate whether the fragments appear at the same relative
position for all centralities or not.  Figure~\ref{fig:NSpecRatioLi}
shows the ratio of the yield of Li to He fragments evaluated at
$\eta'$\,=\,2.02.
The three panels show the same data as a function of (a) $N_{\rm spec}/2$, (b) $N_{\rm part}/2$, and (c) the
collision geometry ($N_{\rm spec}/2A$).
Between Au+Au and Cu+Cu collisions, the Li/$\alpha$ ratios clearly
do not exhibit a scaling with either $N_{\rm part}/2$ 
(i.e. a similar Li/$\alpha$ ratio at a similar $N_{\rm part}/2$)
or with collision geometry.
The collision geometry, defined as $N_{\rm spec}/2A$,
represents the fraction of total nuclear volume
which interacts such that the overlap shape for each nucleus is roughly
similar.  
A scaling with $N_{\rm spec}/2$ is suggested by the data -- the decreased ratio would
indicate that the emission of the lighter fragments is favored for
fewer spectator nucleons from the collision system.
However, the possibility that this ratio for each system is
constant with centrality is not ruled out within the systematic
uncertainty.  For this scenario, the lower Cu+Cu ratio would
indicate a more favorable emission of the lighter fragment in the
Cu+Cu system than in Au+Au collisions.

From this data, one may attempt to draw a picture of the emission process for
fragments.
Unless the spectator nucleons acquire some $p_{T}$ from intrinsic
Fermi motion or the collision process itself, they would simply
travel straight down the beam pipe until the 
magnetic field of the RHIC steering magnets bent them
away.  In such a case, they would not be visible in the
detector as these magnets are located too far from the
apparatus to have had any influence on the fragments.
The movement of the fragments must be connected to the
nucleus and/or be the result of the collision.

In the simplest scenario, the fragments would move outward
due to their intrinsic (precollision) motion, without further
interaction. This, however, would result in the centrality
and pseudorapidity dependencies being decoupled from each other.
Specifically, the data in every pseudorapidity interval should have
the same centrality dependence (although with different yields);
this is not seen in the data.
Figures~\ref{fig:NSpecRatio}~and~\ref{fig:NSpecRatio2} show the
ratio of $\alpha$ 
yields evaluated at $\eta'$\,=\,1.57 and $\eta'$\,=\,1.21,
respectively, divided by the yield at $\eta'$\,=\,2.02,
for both Au+Au and Cu+Cu collision systems.
The three panels show the same data as a function of (a) $N_{\rm spec}/2$,
(b) $N_{\rm part}/2$, and (c) the
collision geometry.

The ratios in Figs.~\ref{fig:NSpecRatio}~and~\ref{fig:NSpecRatio2}
are not constant as the number of $\alpha$ particles in each
$\eta'$ range ($\eta'$\,=\,1.57 and 1.21, respectively) diminishes
(compared to the reference at $\eta'$\,=\,2.02) with decreasing centrality.
Effectively, the $\alpha$ particles are moving out of the acceptance of the
detector for more peripheral collisions and the
average deflection away from the beam direction
increases for more central collisions.
Such a deflection is suggestive of a specific dependence of
transverse momentum acquired by the fragments.
The same effect is also observed in Cu+Cu collisions.
For fragments moved into the acceptance of PHOBOS due to 
intrinsic (precollision) motion, one would expect 
no centrality dependence of these ratios, i.e. all flat.
Comparing the Cu+Cu and Au+Au data in the three scaling scenarios,
it is apparent that these ratios favor a scaling with $N_{\rm part}/2$, which is perhaps counter-intuitive as these spectators are
often considered to be independent of interactions in the hot participant zone.

\section{Conclusion}

In conclusion, nuclear fragments ($Z$\,$>$\,1) have
been observed up to $Z$\,=\,7 using the extensive reach in pseudorapidity
of the PHOBOS detector.  The pseudorapidity and centrality dependence is
shown for fragments up to $Z$\,=\,5 only for Au+Au; for Cu+Cu this study
is restricted to $Z$\,=\,2~and~3.
Fragments from Au+Au
($\snn$\,=\,19.6\,GeV) and Cu+Cu ($\snn$\,=\,22.4\,GeV) collisions
have sufficiently low longitudinal momentum that
even fragments which have a modest \pT~are deflected into
the PHOBOS apparatus.  The yield of $\alpha$ fragments
is observed to be similar to that measured in other
experiments over a range of energies if evaluated at the
same value of $\eta-y_{\rm beam}$.  As a function of centrality,
the yield of $\alpha$ and Lithium fragments is found to approximately scale
with the number of spectators in the collision.
The centrality dependence of ratios of $\alpha$ fragment yields at
different pseudorapidities illustrates that these fragments move
out of the acceptance of the detector for more peripheral collisions.
In comparing Cu+Cu and Au+Au ratios, a scaling with the number
of participants is favored, suggesting an influence of the 
hot participant zone with the nonparticipating spectators.

\appendix

\section{Relating $y$ and $\eta$}\label{ap:EtaPrimer}

Rapidity, $y$, is defined in Eq.~\ref{eq:y} from Ref.~\cite{PDB} and
has a simple one-to-one relationship with the longitudinal
velocity, $\beta_{z}$:
\beq
y \equiv \frac{1}{2} \ln \left( \frac{E+p_z}{E-p_z} \right) = \tanh^{-1}\left( \frac{p_z}{E} \right) = \tanh^{-1}{\beta_z},\label{eq:y}
\eeq

\noindent where $E$ is the total energy of the particle and 
$p_{z}$ is the longitudinal momentum, i.e. the component along the 
beam direction.  In addition, rapidity has the well-known
property that longitudinal boosts are simply additive, where
rapidity differences, $y_1 - y_2$, are invariant under longitudinal
boosts.

In some cases, such as in the PHOBOS multiplicity detector,
only a particle's direction ($\theta$ -- polar angle and $\phi$ --
azimuthal angle) is accessible, and
not the actual momentum.  In such cases we use the pseudorapidity
variable, $\eta$ -- Eq.~\ref{eq:eta}, from Ref.~\cite{PDB}:
\beq
\eta \equiv -\ln (\tan (\theta/2)),\label{eq:eta}
\eeq
where $\theta$ is the polar angle with respect to the beam direction.
In order to relate these two quantities, one can use two identities
from Ref.~\cite{PDB}:
\beq
p_z = m_T \sinh y,
\eeq
where $m_{T}$ is the transverse mass, defined as $m_{T}^{2}$\,=\,$m^{2}+p_{T}^{2}$, and
\beq
p_z = p_T \sinh \eta,
\label{eq:pzvspteta}
\eeq
which can be derived from 
\beq
\sinh \eta = \cot \theta.
\eeq

These identities result in the relation:
\beq
\sinh \eta = (\sinh y) \sqrt{1+\frac{m^2}{p_T^2}}.
\label{eq:master}
\eeq

\subsection*{Mapping $\eta'$ to $y'$ versus $p_{T}/m$.}
The resulting relation between $y$ and $\eta$ (Eq.~\ref{eq:master})
has many implications:

\begin{enumerate}
\item $\eta/y \ge 1$, which leads directly to
\item $y$ and $\eta$ have the same sign, and
\item $|\eta| > |y|$.
\end{enumerate}

One can examine two limits of this relation.
First, in the limit of small $\eta$ (and therefore
also small $y$), $\sinh \eta \rightarrow \eta$ and therefore:
\beq
\eta \approx y \sqrt{1+\frac{m^2}{p_T^2}}.
\eeq

Second, and more importantly for this work, at large $y$
(and therefore also large $\eta$) one can write:
\beq
\sinh y = e^y(1 - e^{-2y})/2 \rightarrow e^y/2.
\label{eq:sinhapprox}
\eeq
Using Eq.~\ref{eq:master} this leads to:
\beq
\eta \approx y + \frac{1}{2} \ln \left(1 +\frac{m^2}{p_T^2} \right ).
\eeq

\noindent Finally, using the definitions: $\eta' \equiv \eta - y_{\rm beam}$ and $y' \equiv y - y_{\rm beam}$:
\beq
\eta' \approx y' + \frac{1}{2} \ln \left( 1 +\frac{m^2}{p_T^2} \right ).
\label{eq:etaprime}
\eeq 

Equation~\ref{eq:etaprime} holds the key information in the relations
between $y'$ and $\eta'$: at large $y$, an $\eta'$ bin corresponds to a
fixed region in $(y',p_T/m)$ space, independent of $y_{\rm beam}$.
Therefore, this formulation represents the best way to compare
$dN/d\eta$ distributions measured at various beam energies.

One can estimate the validity of this approximation by calculating the
absolute error at each rapidity.  An upper bound on the absolute
error from Eq.~\ref{eq:etaprime} is given by
$|\ln (1-e^{-2y})| \approx e^{-2y}$. %graphically verified
For $y$\,$>$\,$2$($>$\,$3$, $>$\,$5$), the error is estimated to
be less than 0.02 ($<$\,2.5$\times10^{-3}$, $<$\,5.0$\times10^{-5}$) units.
Even for $y=1$, the error in the ``large-$y$'' approximation is less than 0.145.

\begin{figure}[!t]
\centering
\includegraphics[angle=0,width=0.475\textwidth]{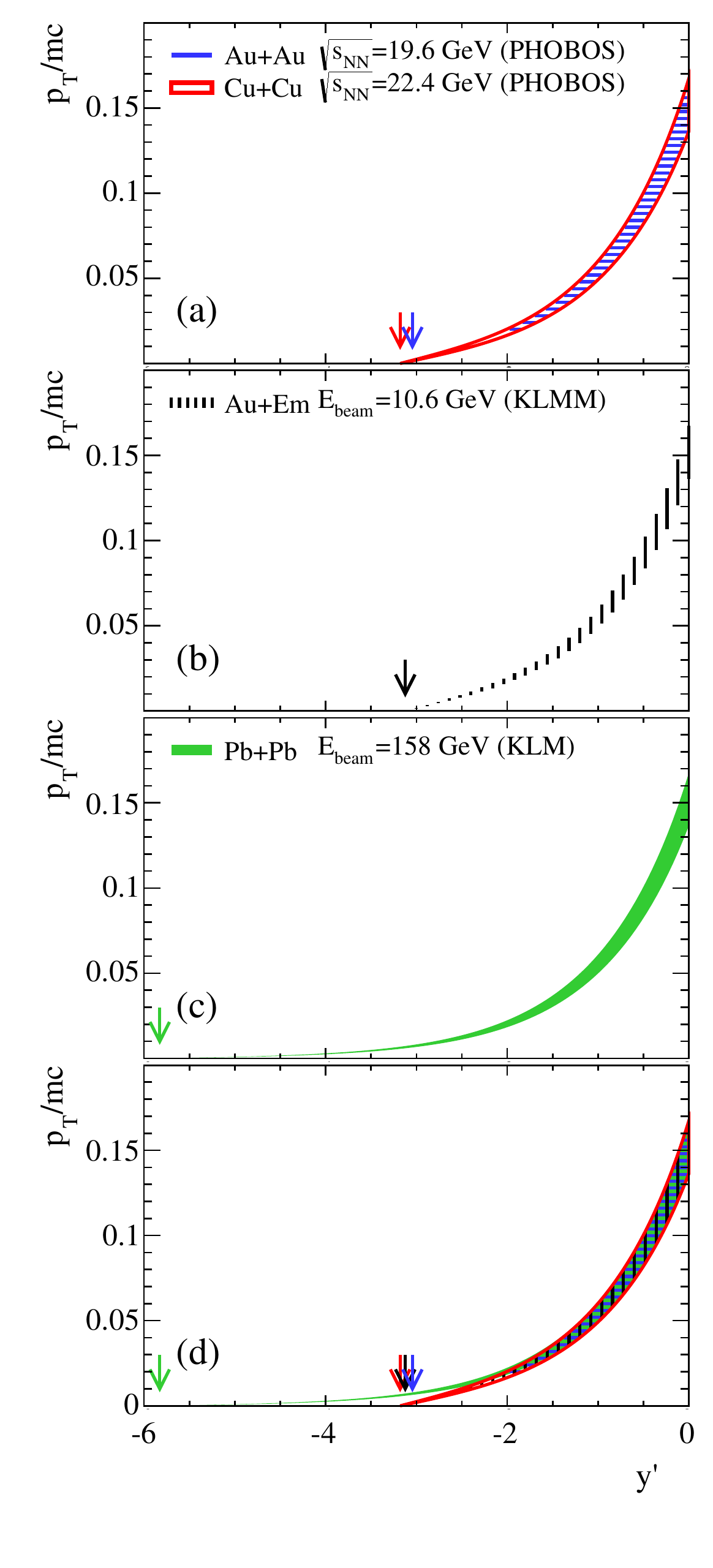}%1
\caption{\label{fig:EtaPrimeAcceptance}
(color online) $p_T/m$-$y'$ acceptance for a fixed $1.8<\eta'<2.0$ window.
The upper (lower) bound on each band corresponds to
$\eta'$\,=\,1.8\,(2.0).  The top three panels (a--c) show the
acceptance for PHOBOS ($\snn$\,=\,19.6\,GeV, and 22.4\,GeV),
KLMM ($E_{\rm beam}$\,=\,10.6\,GeV), and
KLM ($E_{\rm beam}$\,=\,158\,GeV), respectively.
The lower panel (d) shows an overlay of all distributions.
The arrows represent midrapidity ($y$\,=\,0 and $\eta$\,=\,0) at each
energy. See text for discussion.}
\end{figure}

To further illustrate this approximation, for a fixed window in $\eta'$
($1.8<\eta'<2.0$),
Fig.~\ref{fig:EtaPrimeAcceptance} shows the $y'$-$p_T/m$ acceptance.
Panels (a--c) show bands representing the different beam energies used
in this paper: (a) $\snn$\,=\,19.6\,GeV, and 22.4\,GeV representing
Au+Au and Cu+Cu collision data, respectively, measured by PHOBOS,
(b) $E_{\rm beam}$\,=\,10.6\,GeV collisions of Au nuclei on an emulsion
target (Em) measured by KLMM, and (c) $E_{\rm beam}$\,=\,158\,GeV
collisions of Pb nuclei on a stationary Pb target as measured by KLM.
Panel (d) shows an overlay of all distributions.
The arrows represent midrapidity (i.e. $y$\,=\,0 and $\eta$\,=\,0).
The three lowest energy bands (PHOBOS and KLMM) almost entirely
overlap owing to their very similar beam energies (or 
equivalently $y_{\rm beam}$).

In general, to compare results in the rest frame of the beam particle,
PHOBOS has used $\eta'$ to compare pseudorapidity distributions in the
``fragmentation'' or ``extended longitudinal scaling'' region among
data at different energies
($dN_{ch}/d\eta$~\cite{cite:BigMultPaper,cite:SigFrag,cite:PHOBOS_dAuMult,cite:PHOBOS_AuAu62,cite:PHOBOS_CuCu},
and also for the first and second harmonic of the Fourier decomposition of 
the azimuthal angle distribution -- known as 
$v_{1}$~\cite{cite:PHOBOS_v1} and $v_{2}$~\cite{cite:PHOBOS_v2}, respectively).
This is roughly confined to the
$|\eta|>2$ region, so, as shown, $\eta'$ is ideally suited for
this, second only to $y'$ itself.

\subsection*{Limitations}

As Fig.~\ref{fig:EtaPrimeAcceptance} suggests, there are limitations
in this simplification.
There are two important considerations in using $\eta'$ rather than
$y'$.  The first is that the shape in $(y',p_T/m)$ space is
non-intuitive and does {\em not} generally correspond to $\eta'$\,=\,$y'$
except when $p_{T}$\,$\gg$\,$m$.  Therefore, generally interpreting an
$\eta'$ distribution as equivalent to $y'$ can be seriously incorrect
in certain cases.  The second issue is that there can, in principle, be
some contamination to high-$\eta$ from particles with very low $p_{T}$
and $y$ that is not quite beam-energy-independent.  Usually the fact
that these particles would have to come from very low $p_{T}$ helps to
suppress them since the $d^{2}N/dydp_{T}$ yields all go to 0 at
$p_{T}$\,=\,0. In particular, for the region of
$\eta'$$>$0, the mid-rapidity contribution is at {\em particularly} 
low $p_{T}$.
For $\alpha$
particles in this work, the contamination from mid-rapidity can be
expected to be negligible.

When comparing collider data to fixed target data, there is an extra
consideration.  For the positive side $\eta'=\eta-y_{\rm beam}$,
each $\eta'$ bin contains contributions from all positive values of
$y$.  In the case of the collider kinematics this stops at mid-rapidity.
In the case of fixed target kinematics this could, in principle,
include contributions from particles near the target rapidity
(which is 0).  Therefore, some small contamination of $\alpha$
particles emitted at very low $p_T$ from the target rather than
from the $Au$ beam could occur.  Again, this is expected to be negligible,
despite the extent in $\eta$,
since it is at very low $p_T$ and a very narrow window in $p_T$.

\section{Estimation of $dN/dp_{T}$}\label{ap:dNdpt}

The quantity $dN/dp_{T}$ is known to be invariant under
longitudinal boosts and may provide an additional check on 
scaling between data samples at different energies.  The 
measurement of $p_{T}$ is not possible at forward pseudorapidity
in PHOBOS, so an estimate is needed.  It is assumed that the
longitudinal momentum of the spectator nucleons does not change during the
collision.  Given this assumption, one can calculate the transverse momentum as:

\beq
p_{T} = \frac{m~{\rm sinh}(y_{\rm beam})}{{\rm sinh}(\eta)}
\label{eq:pt}
\eeq 

\noindent where $m$ is the mass of the particle of interest ($\alpha$).
Differentiating Eq.~\ref{eq:pt} yields the Jacobian needed to 
transform $dN/d\eta \rightarrow dN/dp_{T}$:

\beq
\frac{d\eta}{dp_{T}} = \frac{d\eta'}{dp_{T}} = -\frac{\rm tanh(\eta)}{p_{T}}
\label{eq:dhdpt}
\eeq 

\begin{figure}[!t]
\centering
\includegraphics[angle=0,width=0.475\textwidth]{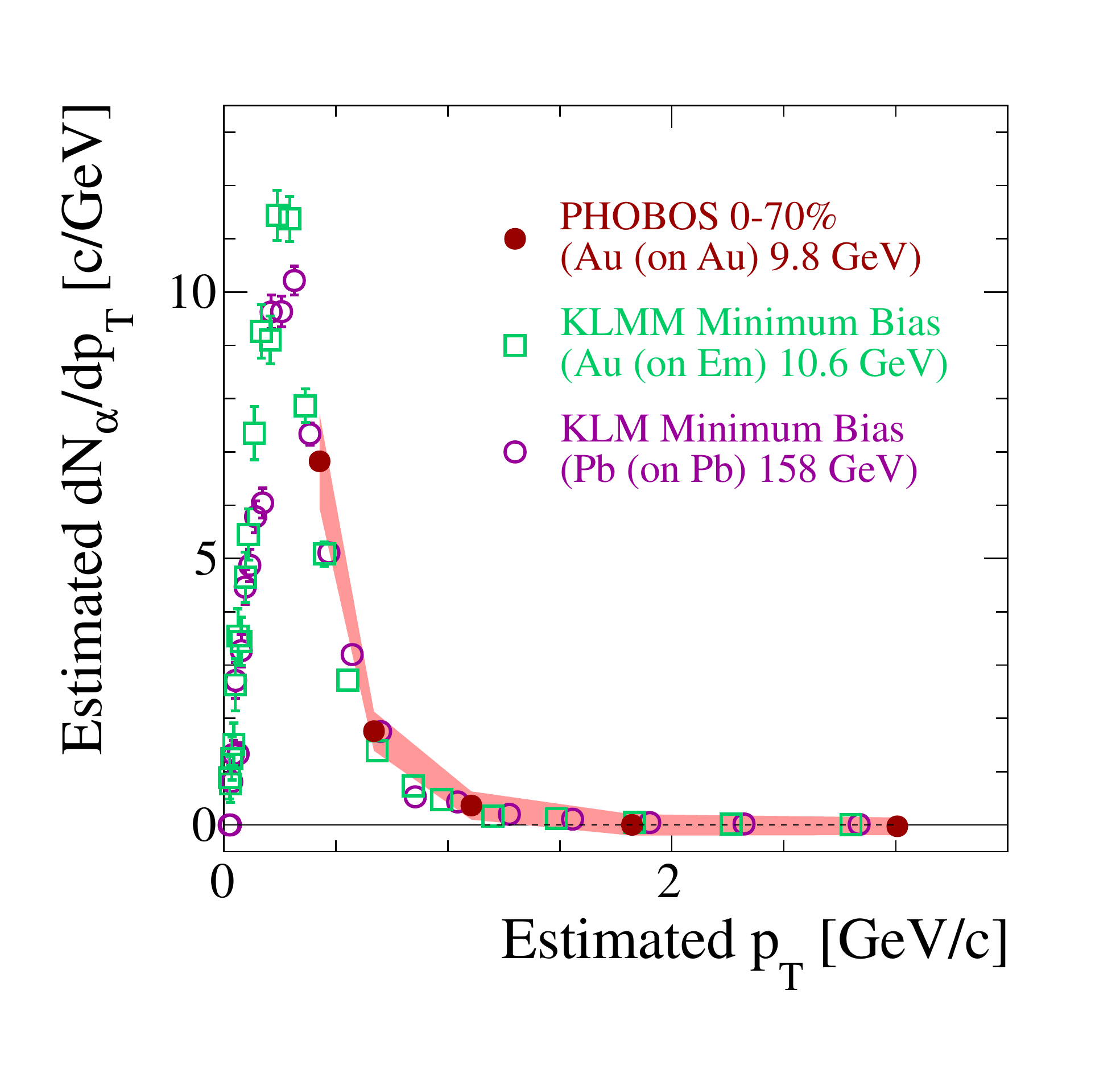}%1
\caption{\label{fig:dNdpt}
  (color online) Estimated $dN/dp_{T}$ distribution for $\alpha$ fragments
  near beam rapidity for 0\%--70\% central Au+Au collisions at $\snn$\,=\,19.6\,GeV.
  Estimation procedure is described in the text.  For comparison, 
  Au+Em ($\snn$\,=\,4.6\,GeV)~\cite{cite:KLMM_PhysRevC}  and
  Pb+Pb ($\snn$\,=\,17.2\,GeV)~\cite{cite:KLM_ActaPhysPol} collisions
  are shown, using the same estimation method.
}
\end{figure}

\begin{figure}[!t]
\centering
\includegraphics[angle=0,width=0.475\textwidth]{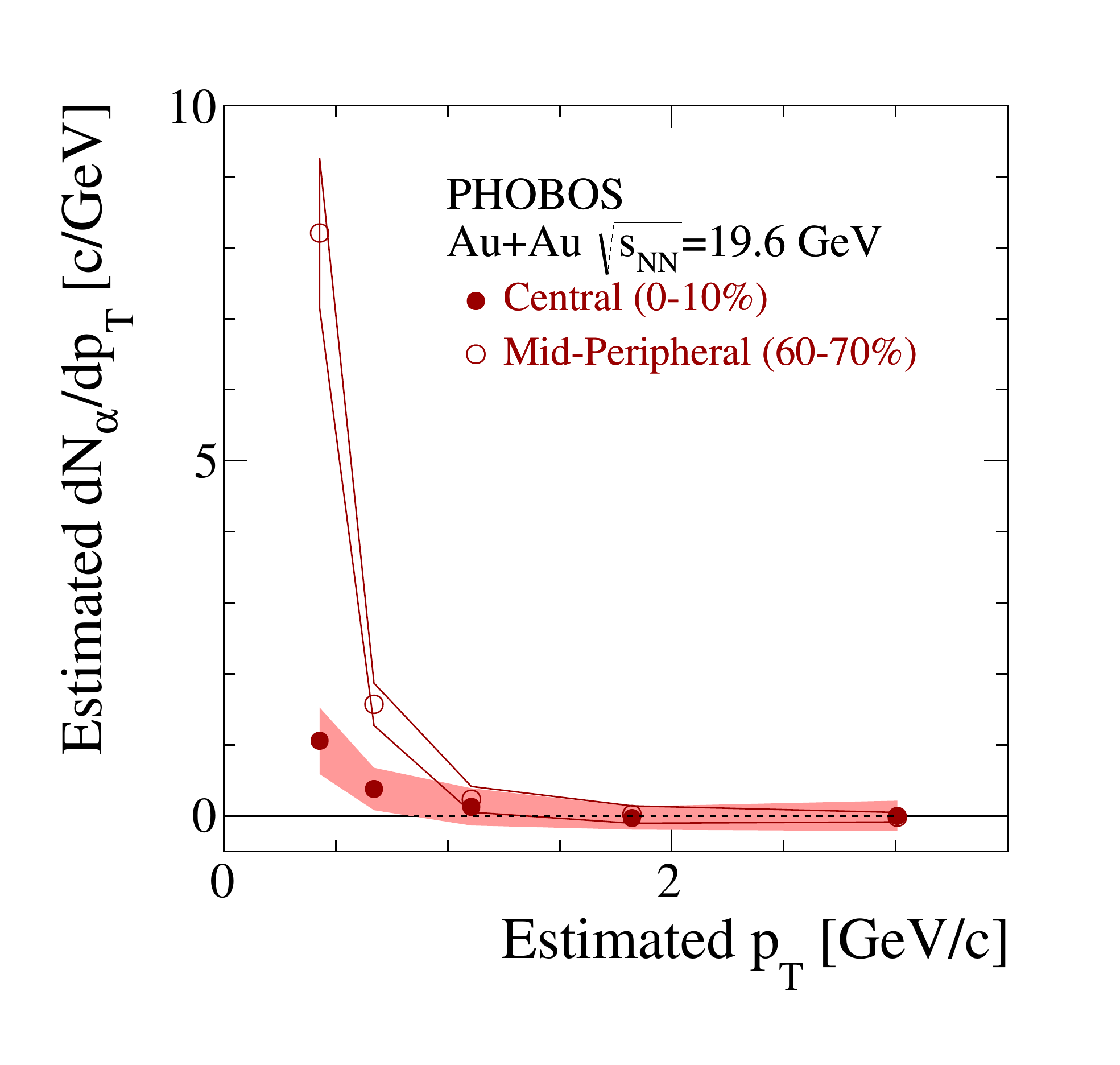}%1
\caption{\label{fig:dNdptCent}
  (color online) Estimated $dN/dp_{T}$ distribution for $\alpha$ fragments 
  near beam rapidity for Au+Au collisions at $\snn$\,=\,19.6\,GeV.  Estimation
  procedure is described in the text.  The open and closed symbols represent 
  central (0\%--10\%) and mid-peripheral (60\%--70\%) collisions, respectively. 
}
\end{figure}

Using these relations (Eq.~\ref{eq:pt}~and~\ref{eq:dhdpt}), one can 
transform $dN/d\eta$ as a function of $\eta$ into $dN/dp_{T}$ as a
function of $p_{T}$.  As a reminder, this is an estimate of both
quantities and is not a precise measurement.  Figure~\ref{fig:dNdpt}
shows a comparison of the estimated $dN/dp_{T}$ versus $p_{T}$ for
0\%--70\% central Au+Au collisions at $\snn$\,=\,19.6\,GeV.  For comparison,
the same technique is used to transform the data from Au+Em
($\snn$\,=\,4.6\,GeV)~\cite{cite:KLMM_PhysRevC}  and
Pb+Pb ($\snn$\,=\,17.2\,GeV)~\cite{cite:KLM_ActaPhysPol} collisions
(i.e. from the data shown in Fig.~\ref{fig:EnergyComparison}).
The data agree well within the uncertainties described above.
Figure~\ref{fig:dNdptCent} shows a comparison between central (closed
symbols) and mid-peripheral (open) Au+Au collisions.
The Cu+Cu data are not shown as the
expected difference in yield between Au (197) fragments
and Cu (63) fragments is large because of the difference in
mass -- whereas the difference
between Au (197) and Pb (208) should be negligible.

\section{Tables of data}\label{ap:Tables}

Table~\ref{tbl:Centrality} shows the $N_{\rm part}$ values determined
from a Glauber model calculation for Au+Au ($\snn$\,=\,19.6\,GeV) and
Cu+Cu ($\snn$\,=\,22.4\,GeV) collisions.

\begin{table}
\centering
  \caption{\label{tbl:Centrality} $N_{\rm part}$ values determined from a Glauber model calculation for Au+Au ($\snn$\,=\,19.6\,GeV) and Cu+Cu ($\snn$\,=\,22.4\,GeV) collisions.  Uncertainties are 90\% C.L. systematic.}
  \begin{tabular}{ccc}
  \hline\hline
  \multirow{2}{*}{\makecell{Centrality\\Bin (\%)}} & \multicolumn{2}{c}{Number of Participants} \\
           &       Au+Au     &      Cu+Cu     \\\hline
     0-10  & 316.3 $\pm$ 9.9 & 93.8 $\pm$ 3.0 \\
    10-20  & 226.5 $\pm$ 8.0 & 68.5 $\pm$ 3.0 \\
    20-30  & 156.5 $\pm$ 7.0 & 48.5 $\pm$ 3.0 \\
    30-40  & 106.0 $\pm$ 7.0 & 33.5 $\pm$ 3.0 \\
    40-50  &  66.0 $\pm$ 4.7 & 22.0 $\pm$ 3.0 \\
    50-60  &  39.5 $\pm$ 3.0 & 14.3 $\pm$ 3.0 \\
    60-70  &  21.3 $\pm$ 3.0 & -- \\
  \hline\hline
  \end{tabular}
\end{table}

Tables~\ref{tbl:AuAuAlphas}--\ref{tbl:AuAuBoron}~and~\ref{tbl:CuCuAlphas}--\ref{tbl:CuCuLithium}
contain the corrected $dN_{\rm particle}/d\eta$ yields as 
function of collision centrality for Au+Au ($\snn$\,=\,19.6\,GeV) 
and Cu+Cu ($\snn$\,=\,22.4\,GeV) collisions, respectively.
Note that for clarity some values are scaled up by powers of 10.

\begin{table*}
\centering
  \caption{\label{tbl:AuAuAlphas} $dN_{\alpha}/d\eta$ measured in Au+Au collisions at $\snn$\,=\,19.6\,GeV.  Uncertainties are 1-$\sigma$ statistical and 90\% C.L. systematic.}
  \begin{tabular}{cccccc}
  \hline\hline
\multirow{2}{*}{\makecell{Centrality\\Bin (\%)}} & \multicolumn{5}{c}{Yield} \\
 & 3.0$<$\,$|\eta|$\,$<$3.5 & 3.5$<$\,$|\eta|$\,$<$4.0 & 4.0$<$\,$|\eta|$\,$<$4.5 & 4.5$<$\,$|\eta|$\,$<$5.0 & 5.0$<$\,$|\eta|$\,$<$5.4 \\\hline
     0-10  &  0.01 $\pm$  0.03 $\pm$  0.65 & -0.05 $\pm$  0.02 $\pm$  0.30 &  0.14 $\pm$  0.01 $\pm$  0.29 &  0.26 $\pm$  0.01 $\pm$  0.20 &  0.45 $\pm$  0.02 $\pm$  0.20 \\ 
    10-20  & -0.00 $\pm$  0.02 $\pm$  0.46 & -0.24 $\pm$  0.02 $\pm$  0.86 &  0.41 $\pm$  0.02 $\pm$  0.36 &  0.84 $\pm$  0.02 $\pm$  0.20 &  1.67 $\pm$  0.03 $\pm$  0.20 \\ 
    20-30  & -0.08 $\pm$  0.02 $\pm$  0.55 &  0.09 $\pm$  0.01 $\pm$  0.33 &  0.52 $\pm$  0.02 $\pm$  0.30 &  1.38 $\pm$  0.02 $\pm$  0.20 &  2.95 $\pm$  0.04 $\pm$  0.27 \\ 
    30-40  & -0.25 $\pm$  0.02 $\pm$  0.64 &  0.12 $\pm$  0.01 $\pm$  0.33 &  0.57 $\pm$  0.02 $\pm$  0.34 &  1.64 $\pm$  0.02 $\pm$  0.28 &  3.82 $\pm$  0.04 $\pm$  0.56 \\ 
    40-50  & -0.22 $\pm$  0.02 $\pm$  0.70 &  0.04 $\pm$  0.01 $\pm$  0.33 &  0.51 $\pm$  0.02 $\pm$  0.29 &  1.68 $\pm$  0.02 $\pm$  0.43 &  4.06 $\pm$  0.04 $\pm$  0.45 \\ 
    50-60  &  0.05 $\pm$  0.01 $\pm$  0.33 &  0.01 $\pm$  0.01 $\pm$  0.20 &  0.40 $\pm$  0.01 $\pm$  0.28 &  1.40 $\pm$  0.02 $\pm$  0.22 &  3.95 $\pm$  0.04 $\pm$  0.52 \\ 
    60-70  & -0.05 $\pm$  0.01 $\pm$  0.20 &  0.04 $\pm$  0.01 $\pm$  0.22 &  0.26 $\pm$  0.01 $\pm$  0.20 &  1.05 $\pm$  0.02 $\pm$  0.20 &  3.51 $\pm$  0.04 $\pm$  0.45 \\ 
  \hline\hline
  \end{tabular}
\end{table*}

\begin{table*}
\centering
  \caption{\label{tbl:AuAuLithium} $dN_{Li}/d\eta$ measured in Au+Au collisions at $\snn$\,=\,19.6\,GeV.  Uncertainties are 1-$\sigma$ statistical and 90\% C.L. systematic.  Yields are scaled up by a factor of 10 for clarity.}
  \begin{tabular}{cccccc}
  \hline\hline
\multirow{2}{*}{\makecell{Centrality\\Bin (\%)}} & \multicolumn{5}{c}{Yield $\times$ 10} \\
 & 3.0$<$\,$|\eta|$\,$<$3.5 & 3.5$<$\,$|\eta|$\,$<$4.0 & 4.0$<$\,$|\eta|$\,$<$4.5 & 4.5$<$\,$|\eta|$\,$<$5.0 & 5.0$<$\,$|\eta|$\,$<$5.4 \\\hline
     0-10  &  0.41 $\pm$  0.07 $\pm$  0.47 &  0.01 $\pm$  0.05 $\pm$  0.34 &  0.05 $\pm$  0.03 $\pm$  0.22 &  0.03 $\pm$  0.03 $\pm$  0.20 &  0.08 $\pm$  0.04 $\pm$  0.51 \\ 
    10-20  &  0.06 $\pm$  0.07 $\pm$  0.48 & -0.19 $\pm$  0.05 $\pm$  0.76 &  0.08 $\pm$  0.04 $\pm$  0.29 &  0.17 $\pm$  0.04 $\pm$  0.20 &  0.66 $\pm$  0.07 $\pm$  0.20 \\ 
    20-30  &  0.12 $\pm$  0.06 $\pm$  0.36 &  0.09 $\pm$  0.04 $\pm$  0.36 &  0.13 $\pm$  0.04 $\pm$  0.33 &  0.26 $\pm$  0.04 $\pm$  0.20 &  1.23 $\pm$  0.09 $\pm$  0.20 \\ 
    30-40  & -0.05 $\pm$  0.06 $\pm$  0.43 &  0.13 $\pm$  0.04 $\pm$  0.24 &  0.02 $\pm$  0.04 $\pm$  0.30 &  0.59 $\pm$  0.05 $\pm$  0.26 &  1.88 $\pm$  0.11 $\pm$  0.63 \\ 
    40-50  & -0.09 $\pm$  0.05 $\pm$  0.51 &  0.01 $\pm$  0.03 $\pm$  0.27 &  0.25 $\pm$  0.04 $\pm$  0.35 &  0.66 $\pm$  0.05 $\pm$  0.37 &  1.88 $\pm$  0.11 $\pm$  0.20 \\ 
    50-60  &  0.03 $\pm$  0.04 $\pm$  0.25 & -0.02 $\pm$  0.03 $\pm$  0.21 &  0.14 $\pm$  0.03 $\pm$  0.20 &  0.41 $\pm$  0.05 $\pm$  0.20 &  1.71 $\pm$  0.11 $\pm$  0.25 \\ 
    60-70  &  0.04 $\pm$  0.03 $\pm$  0.20 &  0.08 $\pm$  0.02 $\pm$  0.28 &  0.02 $\pm$  0.03 $\pm$  0.20 &  0.28 $\pm$  0.04 $\pm$  0.20 &  1.19 $\pm$  0.10 $\pm$  0.30 \\ 
  \hline\hline
  \end{tabular}
\end{table*}

\begin{table*}
\centering
  \caption{\label{tbl:AuAuBeryllium} $dN_{Be}/d\eta$ measured in Au+Au collisions at $\snn$\,=\,19.6\,GeV.  Uncertainties are 1-$\sigma$ statistical and 90\% C.L. systematic. Yields are scaled up by a factor of 100 for clarity.}
  \begin{tabular}{cccccc}
  \hline\hline
\multirow{2}{*}{\makecell{Centrality\\Bin (\%)}} & \multicolumn{5}{c}{Yield $\times$ 100} \\
 & 3.0$<$\,$|\eta|$\,$<$3.5 & 3.5$<$\,$|\eta|$\,$<$4.0 & 4.0$<$\,$|\eta|$\,$<$4.5 & 4.5$<$\,$|\eta|$\,$<$5.0 & 5.0$<$\,$|\eta|$\,$<$5.4 \\\hline
     0-10  & 0.09 $\pm$  0.45 $\pm$  0.73 & -0.43 $\pm$  0.30 $\pm$  1.14 & -0.24 $\pm$  0.22 $\pm$  0.59 & -0.31 $\pm$  0.14 $\pm$  0.60 &  0.26 $\pm$  0.22 $\pm$  0.41 \\ 
    10-20  &  0.02 $\pm$  0.41 $\pm$  0.65 & -0.53 $\pm$  0.30 $\pm$  1.58 & -0.38 $\pm$  0.25 $\pm$  1.02 &  0.26 $\pm$  0.21 $\pm$  0.38 &  1.17 $\pm$  0.36 $\pm$  0.96 \\ 
    20-30  & -0.37 $\pm$  0.37 $\pm$  0.95 & -0.52 $\pm$  0.25 $\pm$  1.18 &  0.18 $\pm$  0.26 $\pm$  0.63 &  0.58 $\pm$  0.27 $\pm$  0.43 &  2.09 $\pm$  0.49 $\pm$  1.24 \\ 
    30-40  & -0.43 $\pm$  0.36 $\pm$  1.46 &  0.30 $\pm$  0.25 $\pm$  0.48 &  0.36 $\pm$  0.25 $\pm$  1.07 & -0.12 $\pm$  0.24 $\pm$  0.43 &  2.84 $\pm$  0.58 $\pm$  0.73 \\ 
    40-50  & -0.47 $\pm$  0.29 $\pm$  1.22 & -0.25 $\pm$  0.21 $\pm$  0.73 & -0.46 $\pm$  0.22 $\pm$  0.83 &  0.55 $\pm$  0.26 $\pm$  0.52 &  3.13 $\pm$  0.61 $\pm$  0.66 \\ 
    50-60  & -0.14 $\pm$  0.23 $\pm$  0.59 &  0.21 $\pm$  0.19 $\pm$  0.34 &  0.13 $\pm$  0.23 $\pm$  0.37 &  0.53 $\pm$  0.26 $\pm$  0.47 &  2.29 $\pm$  0.55 $\pm$  2.32 \\ 
    60-70  & -0.24 $\pm$  0.17 $\pm$  0.42 &  0.03 $\pm$  0.15 $\pm$  0.26 & -0.20 $\pm$  0.18 $\pm$  0.66 &  0.24 $\pm$  0.23 $\pm$  0.47 &  2.04 $\pm$  0.51 $\pm$  1.32 \\ 
  \hline\hline
  \end{tabular}
\end{table*}

\begin{table*}
\centering
  \caption{\label{tbl:AuAuBoron} $dN_{B}/d\eta$ measured in Au+Au collisions at $\snn$\,=\,19.6\,GeV.  Uncertainties are 1-$\sigma$ statistical and 90\% C.L. systematic.  Yields are scaled up by a factor of 100 for clarity.}
  \begin{tabular}{cccccc}
  \hline\hline
\multirow{2}{*}{\makecell{Centrality\\Bin (\%)}} & \multicolumn{5}{c}{Yield $\times$ 100} \\
 & 3.0$<$\,$|\eta|$\,$<$3.5 & 3.5$<$\,$|\eta|$\,$<$4.0 & 4.0$<$\,$|\eta|$\,$<$4.5 & 4.5$<$\,$|\eta|$\,$<$5.0 & 5.0$<$\,$|\eta|$\,$<$5.4 \\\hline
     0-10  & -0.09 $\pm$  0.41 $\pm$  1.15 & -0.19 $\pm$  0.29 $\pm$  0.75 & -0.04 $\pm$  0.23 $\pm$  0.84 & -0.21 $\pm$  0.16 $\pm$  0.60 & -0.29 $\pm$  0.10 $\pm$  0.63 \\ 
    10-20  &  0.17 $\pm$  0.40 $\pm$  0.80 &  0.70 $\pm$  0.35 $\pm$  1.04 &  0.41 $\pm$  0.28 $\pm$  1.03 &  0.06 $\pm$  0.18 $\pm$  0.43 &  0.32 $\pm$  0.29 $\pm$  0.37 \\ 
    20-30  & -0.45 $\pm$  0.36 $\pm$  1.01 & -0.45 $\pm$  0.24 $\pm$  1.34 & -0.39 $\pm$  0.23 $\pm$  1.43 & -0.23 $\pm$  0.22 $\pm$  0.43 &  1.60 $\pm$  0.46 $\pm$  0.94 \\ 
    30-40  &  0.38 $\pm$  0.40 $\pm$  0.95 &  0.17 $\pm$  0.26 $\pm$  0.76 &  0.16 $\pm$  0.25 $\pm$  1.01 &  0.01 $\pm$  0.24 $\pm$  0.54 &  2.31 $\pm$  0.56 $\pm$  1.03 \\ 
    40-50  & 0.45 $\pm$  0.31 $\pm$  0.95 &  0.05 $\pm$  0.22 $\pm$  0.82 & -0.37 $\pm$  0.22 $\pm$  1.30 &  0.05 $\pm$  0.23 $\pm$  0.78 &  2.01 $\pm$  0.56 $\pm$  0.86 \\ 
    50-60  & -0.11 $\pm$  0.22 $\pm$  0.89 &  0.07 $\pm$  0.18 $\pm$  0.50 & -0.02 $\pm$  0.22 $\pm$  0.58 &  0.34 $\pm$  0.25 $\pm$  0.50 &  3.36 $\pm$  0.57 $\pm$  1.29 \\ 
    60-70  & -0.29 $\pm$  0.15 $\pm$  0.75 & -0.04 $\pm$  0.15 $\pm$  0.68 & -0.11 $\pm$  0.18 $\pm$  0.47 &  0.03 $\pm$  0.21 $\pm$  0.38 &  1.71 $\pm$  0.48 $\pm$  0.46 \\ 
  \hline\hline
  \end{tabular}
\end{table*}

\begin{table*}
\centering
  \caption{\label{tbl:CuCuAlphas} $dN_{\alpha}/d\eta$ measured in Cu+Cu collisions at $\snn$\,=\,22.4\,GeV.  Uncertainties are 1-$\sigma$ statistical and 90\% C.L. systematic. Yields are scaled up by a factor of 10 for clarity.}
  \begin{tabular}{cccccc}
  \hline\hline
\multirow{2}{*}{\makecell{Centrality\\Bin (\%)}} & \multicolumn{5}{c}{Yield $\times$ 10} \\
 & 3.0$<$\,$|\eta|$\,$<$3.5 & 3.5$<$\,$|\eta|$\,$<$4.0 & 4.0$<$\,$|\eta|$\,$<$4.5 & 4.5$<$\,$|\eta|$\,$<$5.0 & 5.0$<$\,$|\eta|$\,$<$5.4 \\\hline
     0-10  & -0.15 $\pm$  0.16 $\pm$  0.50 & -0.16 $\pm$  0.09 $\pm$  0.56 &  0.28 $\pm$  0.04 $\pm$  0.50 &  0.87 $\pm$  0.07 $\pm$  0.50 &  1.94 $\pm$  0.07 $\pm$  0.50 \\ 
    10-20  &  0.21 $\pm$  0.08 $\pm$  0.50 &  0.21 $\pm$  0.14 $\pm$  0.50 &  0.63 $\pm$  0.06 $\pm$  0.50 &  2.04 $\pm$  0.09 $\pm$  0.50 &  5.06 $\pm$  0.09 $\pm$  0.79 \\ 
    20-30  &  0.11 $\pm$  0.09 $\pm$  0.50 &  0.20 $\pm$  0.08 $\pm$  0.50 &  0.77 $\pm$  0.06 $\pm$  0.50 &  2.62 $\pm$  0.10 $\pm$  0.55 &  7.35 $\pm$  0.12 $\pm$  1.34 \\ 
    30-40  &  0.08 $\pm$  0.08 $\pm$  0.50 &  0.09 $\pm$  0.07 $\pm$  0.50 &  0.75 $\pm$  0.06 $\pm$  0.50 &  2.80 $\pm$  0.07 $\pm$  0.50 &  8.09 $\pm$  0.12 $\pm$  1.37 \\ 
    40-50  &  0.13 $\pm$  0.10 $\pm$  0.50 &  0.11 $\pm$  0.06 $\pm$  0.50 &  0.66 $\pm$  0.05 $\pm$  0.50 &  2.47 $\pm$  0.06 $\pm$  0.50 &  7.23 $\pm$  0.11 $\pm$  1.19 \\ 
    50-60  &  0.08 $\pm$  0.05 $\pm$  0.50 &  0.12 $\pm$  0.06 $\pm$  0.50 &  0.45 $\pm$  0.04 $\pm$  0.50 &  1.95 $\pm$  0.05 $\pm$  0.50 &  5.99 $\pm$  0.10 $\pm$  1.04 \\ 
  \hline\hline
  \end{tabular}
\end{table*}

\begin{table*}
\centering
  \caption{\label{tbl:CuCuLithium} $dN_{Li}/d\eta$ measured in Cu+Cu collisions at $\snn$\,=\,22.4\,GeV.  Uncertainties are 1-$\sigma$ statistical and 90\% C.L. systematic. Yields are scaled up by a factor of 100 for clarity.}
  \begin{tabular}{cccccc}
  \hline\hline
\multirow{2}{*}{\makecell{Centrality\\Bin (\%)}} & \multicolumn{5}{c}{Yield $\times$ 100} \\
 & 3.0$<$\,$|\eta|$\,$<$3.5 & 3.5$<$\,$|\eta|$\,$<$4.0 & 4.0$<$\,$|\eta|$\,$<$4.5 & 4.5$<$\,$|\eta|$\,$<$5.0 & 5.0$<$\,$|\eta|$\,$<$5.4 \\\hline
     0-10  &  0.61 $\pm$  0.40 $\pm$  2.40 &  0.41 $\pm$  0.37 $\pm$  0.71 &  0.25 $\pm$  0.09 $\pm$  0.53 &  0.37 $\pm$  0.08 $\pm$  0.40 &  0.03 $\pm$  0.12 $\pm$  0.42 \\ 
    10-20  &  0.05 $\pm$  0.22 $\pm$  2.15 &  0.12 $\pm$  0.22 $\pm$  0.58 &  0.39 $\pm$  0.10 $\pm$  0.80 &  0.46 $\pm$  0.10 $\pm$  0.54 &  0.40 $\pm$  0.17 $\pm$  0.40 \\ 
    20-30  &  0.45 $\pm$  0.18 $\pm$  1.03 &  0.14 $\pm$  0.20 $\pm$  0.77 &  0.19 $\pm$  0.09 $\pm$  0.61 &  0.13 $\pm$  0.12 $\pm$  0.58 &  0.96 $\pm$  0.23 $\pm$  0.49 \\ 
    30-40  &  0.44 $\pm$  0.18 $\pm$  1.25 &  0.28 $\pm$  0.18 $\pm$  0.68 &  0.24 $\pm$  0.09 $\pm$  0.40 &  0.19 $\pm$  0.08 $\pm$  0.66 &  1.23 $\pm$  0.31 $\pm$  0.68 \\ 
    40-50  &  0.28 $\pm$  0.11 $\pm$  0.80 &  0.26 $\pm$  0.15 $\pm$  0.41 &  0.14 $\pm$  0.07 $\pm$  0.40 &  0.17 $\pm$  0.08 $\pm$  0.40 &  1.22 $\pm$  0.22 $\pm$  0.40 \\ 
    50-60  &  0.20 $\pm$  0.10 $\pm$  0.65 &  0.20 $\pm$  0.10 $\pm$  0.40 &  0.03 $\pm$  0.07 $\pm$  0.40 &  0.10 $\pm$  0.07 $\pm$  0.40 &  0.90 $\pm$  0.24 $\pm$  0.41 \\ 
  \hline\hline
  \end{tabular}
\end{table*}

\begin{acknowledgments}
This work was partially supported by U.S. DOE grants
DE-AC02-98CH10886, DE-FG02-93ER40802,
DE-FG02-94ER40818, DE-FG02-94ER40865,
DE-FG02-99ER41099, and DE-AC02-06CH11357,
by U.S. NSF grants
9603486, 0072204, and 0245011,
%by Polish MNiSW grant N N202 282234 (2008-2010),
by Polish National Science Center grant DEC-2013/08/M/ST2/00320,
by NSC of Taiwan Contract
NSC 89-2112-M-008-024,
and by
Hungarian OTKA grant
(F 049823).
\end{acknowledgments}

\end{document}